\documentclass[english,nofootinbib]{revtex4-2}
\usepackage[T1]{fontenc}
\usepackage[latin9]{inputenc}
\usepackage{color}
\usepackage{array}
\usepackage{float}
\usepackage{units}
\usepackage{mathrsfs}
\usepackage{mathtools}
\usepackage{amsmath}
\usepackage{stackrel}
\usepackage{graphicx}

\makeatletter

\newcommand*\LyXThinSpace{\,\hspace{0pt}}
\providecommand{\tabularnewline}{\\}

\usepackage{babel}

\makeatother

\usepackage{babel}
\begin{document}
\title{New protocols for quantum key distribution with explicit upper and
lower bound on secret-key rate}
\author{Arindam Dutta}
\email{arindamsalt@gmail.com}

\email{https://orcid.org/0000-0003-3909-7519}

\author{Anirban Pathak}
\email{anirban.pathak@gmail.com}

\email{https://orcid.org/0000-0003-4195-2588}

\affiliation{Department of Physics and Materials Science \& Engineering, Jaypee
Institute of Information Technology, A 10, Sector 62, Noida, UP-201309,
India}
\begin{abstract}
We present two new schemes for quantum key distribution (QKD) that
neither require entanglement nor an ideal single photon source, making
them implementable with commercially available single photon sources.
These protocols are shown to be secure against multiple attacks, including
intercept-resend and a class of collective attacks. We derive bounds
on the key rate and demonstrate that a specific type of classical
pre-processing can increase the tolerable error limit. A trade-off
between quantum resources and information revealed to an eavesdropper
(Eve) is observed, with higher efficiency achievable through the use
of additional quantum resources. Specifically, our proposed protocols
outperform the SARG04 protocol in terms of efficiency at the cost
of more quantum resources.
\end{abstract}
\maketitle

\section{introduction}

Cryptography has been an essential and useful technique for mankind
since the beginning of civilization. Historically, cryptographic methods
have been employed to camouflage secret information, but cryptanalysts
often develop more powerful methods to decipher these secret messages.
A paradigm shift in cryptography occurred in the 1970s with the introduction
of public key cryptography methods, such as RSA \cite{RSA78} and
Diffie-Hellman (DH) \cite{DH76} schemes. The security of these and
other classical key distribution schemes arises from the complexity
of the computational tasks inherently used in their design. For example,
the security of the RSA scheme and DH scheme relies on the computational
complexity of the factorization of an odd bi-prime problem and the
discrete logarithm problem, respectively \cite{P13}.

In a seminal work in 1994, Peter W. Shor \cite{S94} demonstrated
that both the factorization of an odd bi-prime problem and the discrete
logarithm problem could be solved efficiently (i.e., in polynomial
time) using quantum computers. This finding implies that many conventional
key distribution methods would be vulnerable if a scalable quantum
computer were developed. Thus, cryptography faces a significant challenge
from quantum computers or, more precisely, from quantum algorithms
that can solve several computational tasks much faster than their
classical counterparts. Interestingly, a solution to the challenge
posed by quantum computers already exists through quantum key distribution
(QKD) methods. In QKD, key distribution is facilitated by quantum
resources, and security is derived from the fundamental laws of physics
rather than the computational complexity of a problem. In fact, the
first such scheme for QKD was proposed 10 years before the work of
Shor that put classical cryptography in crisis. The first QKD scheme
was proposed in 1984 by Bennett and Brassard \cite{BB84}. Physical
principles, such as the no-cloning theorem \cite{WZ82}, the collapse
on measurement postulate, and Heisenberg's uncertainty principle,
play crucial roles in establishing the security of this single-qubit-based
scheme, which can be realized using polarization-encoded single photons
and other alternative realizations of photonic qubits. It is important
to note that, ideally, any eavesdropping effort leaves a detectable
trace in a QKD protocol. However, in realistic situations, due to
device imperfections, eavesdropping may occur without causing a detectable
disturbance.

The BB84 protocol was followed by several protocols for QKD \cite{B92,E91,GV95,STP20}
and other related cryptographic tasks \cite{YSP14,SKB+13,LXS_18,BST+17,STP17,thapliyal2018orthogonal,TP15,STP20,DP+23}
(for a review see \cite{SPR17,GRT+02}). Each of these protocols has
its own advantages and disadvantages. Most of these schemes are unconditionally
secure in the ideal situation\footnote{Quantum identity authentication \cite{CS_01,DP22,DP23,DP+24} plays a crucial
role before the execution of a QKD protocol to secure the entire communication.}. However, in real-life situations, the devices used are not perfect,
which leads to side channels for performing quantum hacking using
device imperfections. For example, the BB84 protocol (and many other
protocols of similar nature, like the B92 protocol \cite{B92}) ideally
requires a single photon source, as implementing these types of protocols
necessitates that Alice must be able to send single photon states
to Bob. Currently, commendable experimental efforts have been devoted
to constructing a reliable single-photon source (see \cite{LP21,TS21}
and references therein). However, in most commercial products, weak
coherent pulses (WCPs) produced by attenuating the output of lasers
are used as an approximate single-photon source. The quantum state
of a WCP produced by attenuating a laser can be described as 
\begin{equation}
|\alpha\rangle=|\sqrt{\mu}\exp(i\theta)\rangle=\sum_{n=0}^{\infty}\left(\frac{e^{-\mu}\mu^{n}}{n!}\right)^{\frac{1}{2}}\exp(in\theta)|n\rangle,\label{eq:coherent}
\end{equation}
where $|n\rangle$ represents a Fock state (or equivalently an $n$
photon state) and the mean photon number $\mu=|\alpha|^{2}\ll1$,
Alice produces a quantum state that can be viewed as a superposition
of Fock states with a Poissonian photon number distribution given
by $p(n,\mu)=\frac{e^{-\mu}\mu^{n}}{n!}$. Thus, if such a source
is used, Alice produces the desired one-photon state with a probability
$p(1,\mu)$ and multi-photon pulses with a total probability of $1-p(0,\mu)-p(1,\mu)$.
In this scenario, Alice creates a multi-photon state with the same
information, opening a window that allows Eve to perform a photon
number splitting (PNS) attack \cite{HIG+95}. Further, in long-distance
communication, channel loss is a concern as it allows an eavesdropper
with superior technology to replace the lossy channel with a perfectly
transparent one and perform an eavesdropping attack \cite{BLM+2000},
making it appear as though the effects are due to channel loss. To
counter this, Scarani et al. proposed a QKD scheme (SARG04) in 2004
\cite{SAR+04}, which is robust against PNS attacks. Here, we aim
to propose a set of two new protocols for QKD that would be robust
against PNS attacks (like SARG04) and a family of other attacks, with
some specific advantages over SARG04 and other existing protocols
for QKD with a similar structure.

In every QKD protocol, information splitting occurs. In protocols
like BB84 \cite{BB84} and B92 \cite{B92}, the information is divided
into a classical piece (information about the basis in which the transmitted
qubits are prepared) and a quantum piece (transmitted qubits). A similar
type of information splitting happens in the SARG04 protocol \cite{SAR+04}.
However, in some other protocols, like the Goldenberg-Vaidman (GV)
protocol \cite{GV95}, information is split into two quantum pieces.
The security of all these protocols arises from Eve's inability to
simultaneously access these two or more pieces of information. We
wish to study a foundationally important question that arises from
this observation: Can we modify the efficiency of a protocol and/or
the bounds on the secret-key rate of the protocol by changing it so
that the information contained in the classical piece is reduced?
We will use the SARG04 protocol as our test bed to answer this question.
Specifically, we will introduce two new protocols for QKD that are
similar to the SARG04 protocol but with less information content in
the classical pieces compared to SARG04. The SARG04 protocol was designed
to make the PNS attack \cite{HIG+95} highly improbable but was less
efficient\footnote{Efficiency is computed using Cabello's definition \cite{C2000}. In
this approach, the cost of transferring qubits is the same as the
cost of transferring classical bits, and the quantum channel is not
too noisy, which is not always realistic for long-distance communication
using present technology.} compared to a set of other single-photon-based schemes for QKD. These
facts motivated us to investigate the possibility of overcoming the
PNS attack by leveraging a relatively greater amount of quantum resources
instead of classical ones, with the goal of negating present technological
limitations (e.g., channel loss, channel noise). Specifically, we
aim to propose two new protocols for QKD that will be more efficient
than SARG04 while remaining robust against PNS attacks and other well-known
attacks.

The rest of the paper is organized as follows. In Section \ref{sec:II},
we propose a new single-photon-based protocol for QKD that does not
require an ideal single-photon source. This protocol, referred to
as Protocol 1, is described first in a generalized manner and then
in a step-wise manner. It is shown that a simple modification in the
sifting subprotocol of Protocol 1 leads to a new protocol (Protocol
2) with higher efficiency. A detailed security analysis is provided
in Section \ref{sec:III}. To perform this analysis, we use a depolarizing
channel to represent the error introduced by Eve (or the channel itself),
allowing us to calculate the tolerable error limit for the first quantum
particle sequence prepared by Bob. Additionally, we consider security
against a set of collective attack scenarios. In Section \ref{sec:IV},
we analyze the PNS attack on Protocol 1 and Protocol 2 and calculate
the critical distance, justifying the advantage of using a relatively
higher amount of quantum resources. The paper concludes in Section
\ref{sec:V}.

\section{Proposed QKD protocols\protect\label{sec:II}}

We have previously discussed that many QKD methods involving non-orthogonal
state sequences necessitate dividing information into quantum and
classical components. This division compels Eve to leave traces of
her attempted eavesdropping through measurements. In all such QKD
schemes, Alice and Bob compare the initial state (or basis) prepared
by Alice/Bob and the state received through measurement by Bob/Alice.
This comparison is conducted to detect correlations that may reveal
eavesdropping attempts. Following this step, Alice and Bob retain
the states that meet specific criteria, paving the way for the final
key generation. This stage is often referred to as a classical key-sifting
subprotocol. In this work, we use a bi-directional quantum channel
to distribute quantum information in the form of single photons to
distribute a secret key between two legitimate parties, Alice and
Bob, after the key-sifting subprotocol. Here, Alice has prior information
about the quantum states of her initial sequence that she prepares
to send to Bob. This prior information helps her agree on the position
of the sifted key after information reconciliation.

We assume the following notation: To encode the bit value $x$, Alice
generates the quantum state $\psi_{J}^{x}$, for different encoding
using mutually unbiased bases (MUBs) in a Hilbert space $\mathcal{H}$
of dimension\footnote{Let us suppose two orthonormal bases set in the $d$-dimensional Hilbert
space are $\psi_{j_{1}}:=\{\psi_{1},\psi_{2,}\ldots,\psi_{d}\}$ and
$\psi_{j_{2}}:=\{\psi_{1}^{'},\psi_{2,}^{'}\ldots,\psi_{d}^{'}\},$.
They are called mutually unbiased bases when the square of the magnitude
of the inner product between two different basis elements equals the
inverse of the dimension $d$. This can be expressed as $\left|\langle\psi_{a}|\psi_{b}^{'}\rangle\right|^{2}=\frac{1}{d}$,
$\forall a,b\in\{1,2,\ldots,d\}.$ If one measures the system that
is prepared in one of the MUBs, then the measurement outcome using
another basis will be equally probable or maximally uncertain. } $d$, where $x$ represents the bit value and $J$ represents the
basis used for encoding the bit value $x$. Without loss of generality,
we choose $J\coloneqq\{Z,X\}$, where the basis sets $Z$ and $X$
correspond to $\left\{ |0\rangle,|1\rangle\right\} $ and $\left\{ |+\rangle,|-\rangle\right\} ,$
respectively. The basis sets $Z$ and $X$ are often referred to as
the computational and diagonal basis sets, respectively. For the convenience
of classical key-sifting, we use $J=0$ for $Z$ basis and $J=1$
for the $X$ basis.

Now, using the above notation, we may propose the basic structure
of our protocol in a generalized form as follows:

(1) \emph{ State generation-transmission and measurement:} Alice prepares
and sends a sequence ($S_{A}$) of qubits to Bob, which consists of
one of the four quantum states $\psi_{J}^{x}\coloneqq\{\text{\ensuremath{\psi_{Z}^{x},}}\psi_{X}^{x}\}$
to encode a random sequence of bit value $x\in\{0,1\}$. Bob measures
randomly with computational or diagonal basis and gets a sequence
with one of the three quantum states $\psi_{J}^{y}\coloneqq\{\text{\ensuremath{\psi_{Z}^{x},}}\psi_{X}^{x/x^{\perp}}\}$,
where $\psi_{J}^{x^{\perp}}$ is a state orthogonal to $\psi_{J}^{x}$,
with value $x,y\in\{0,1\}$. At present, we assume that the qubits
being transmitted have not experienced any decoherence. Additionally,
Alice will refrain from sharing basis information with Bob. Up to
this point, the protocol closely resembles the BB84 protocol \cite{BB84}.

Bob generates a sequence ($S_{B1}$) of quantum states based on his
measurement results in $\psi_{J}^{y}$ and sends it to Alice. For
each qubit in the sequence ($S_{B1}$), Alice uses the same basis
as in sequence $S_{A}$ to measure and record the outcome. Alice will
then obtain the state $\psi_{J}^{x^{\perp}}$ with a probability of
$\frac{1}{4}$, given our assumption that the sequence is very long
and the quantum channel is noiseless. If the probability of obtaining
the state $\psi_{J}^{x^{\perp}}$ is within the tolerable (threshold)
limit around $\frac{1}{4}$, Alice will publicly request Bob to transmit
the subsequent qubit sequence, denoted as $(S_{B2})$.

\emph{Preparation and measurement of second sequence }(\emph{$S_{B2}$})\emph{.
}After receiving Alice's request, Bob uses the other MUB (i.e., if
the $Z$ ($X$) basis was used earlier to prepare the $n$th qubit
of the sequence $S_{B1}$, then the $X$ ($Z$) basis will be used
to prepare the $n^{th}$ qubit of the sequence $S_{B2}$) to prepare
the elements of the sequence $S_{B2}$ with the same bit value for
the corresponding positions of the elements ($\psi_{J}^{y}$) of the
sequence $S_{B1}$. Bob sends the sequence $S_{B2}$ to Alice. Alice
measures the received qubits of the sequence $S_{B2}$ using the following
rule: If Alice gets the same state $\psi_{J}^{x}$ after measuring
the qubit sequence $S_{B1}$, she uses the other MUB (second basis).
However, if she gets the state $\psi_{J}^{x^{\perp}}$ (orthogonal
to the corresponding elements of the initial sequence $S_{A}$), she
uses the same basis.

(2) \emph{Condition for key-sifting. }To maximize the fraction of
raw key after the sifting process, we propose a classical subprotocol
that discloses less classical information compared to the SARG04 protocol.
Alice reveals the positions of the qubits for which Bob will retain
the measurement outcomes associated with the elements of the sequence
$S_{A}$ to establish the secret key, subject to two specific conditions:
(a) If Alice obtains orthogonal state $(\psi_{Z}^{x^{\perp}})$ corresponding
to the elements of her initial sequence $(S_{A})$ after measuring
$S_{B1}$, and the measurement result of the sequence $S_{B2}$ is
$\psi_{Z}^{x/x^{\perp}}$, Alice decodes that the Bob's measured state
of sequence $S_{A}$ was $\psi_{X}^{x/x^{\perp}}$. (b) If Alice obtains
the same state $(\psi_{Z}^{x})$ corresponding to the elements of
the sequence $S_{A}$ after measuring $S_{B1}$, and the measurement
result of the second sequence sent by Bob $(S_{B2})$ is $\psi_{X}^{x}$,
then Alice concludes that the measurement result of sequence $S_{A}$
by Bob was $\psi_{Z}^{x}$ if and only if the $J$ value announced
by Bob for the measurement of each element is the same as the $J$
value for the corresponding elements of Alice's initial sequence $S_{A}$.
The $J$ value is revealed only for a subset of qubits. Specifically,
it is disclosed for qubits where Alice's measurement on $S_{B1}$
matches the corresponding elements of $S_{A}$, provided the corresponding
qubits in $S_{B2}$ have matching bit elements in a different basis
of $S_{B1}$.

In the following sections, we will start by providing a detailed step-by-step
explanation of our primary protocol, referred to as Protocol 1. Subsequently,
we will illustrate how a slight modification to the key-sifting subprotocol
within Protocol 1 can enhance the efficiency of our proposed QKD protocol.
This modified version will be denoted as Protocol 2 (please refer
to Table \ref{tab:Main=000020table=000020for=000020protocol=0000201=000020and=0000202}
for more information).

\begin{table}[H]
\caption{\protect\label{tab:Main=000020table=000020for=000020protocol=0000201=000020and=0000202}This
table describes encoding and decoding rules for Protocol 1 and Protocol
2. It also expresses the measurement outcome after the classical sifting
subprotocol.}

\centering{}%
\begin{tabular}{|>{\centering}p{1cm}|>{\centering}p{1cm}|>{\centering}p{1cm}|>{\centering}p{2cm}|>{\centering}p{2cm}|>{\centering}p{1.5cm}|>{\centering}p{1.5cm}|>{\centering}p{1.5cm}|>{\centering}p{1.5cm}|>{\centering}p{1.5cm}|}
\hline 
$S_{A}$  & $S_{B1}$  & $S_{B2}$  & Measurement result of $S_{B1}$ by Alice  & Measurement result of $S_{B2}$ by Alice  & Probability  & $J$ value for P1  & Result determine by P1  & $M$ value for P2  & Result determine by P2\tabularnewline
\hline 
 & $|0\rangle$  & $|+\rangle$  & $|0\rangle$  & $|+\rangle$  & $\nicefrac{1}{8}$  & 0  & $|0\rangle$  & 0  & $|0\rangle$\tabularnewline
\cline{2-10}
 &  &  & $|0\rangle$  & $|+\rangle$  & $\nicefrac{1}{64}$  & 1  & $-$  & 0  & $|0\rangle$\tabularnewline
 & $|+\rangle$  & $|0\rangle$  & $|0\rangle$  & $|-\rangle$  & $\nicefrac{1}{64}$  & $-$  & $-$  & 0  & $|+\rangle$\tabularnewline
$|0\rangle$  &  &  & $|1\rangle$  & $|0\rangle$  & $\nicefrac{1}{32}$  & $-$  & $|+\rangle$  & $-$  & $|+\rangle$\tabularnewline
\cline{2-10}
 &  &  & $|0\rangle$  & $|+\rangle$  & $\nicefrac{1}{64}$  & 1  & $-$  & 1  & $|-\rangle$\tabularnewline
 & $|-\rangle$  & $|1\rangle$  & $|0\rangle$  & $|-\rangle$  & $\nicefrac{1}{64}$  & $-$  & $-$  & 1  & $|-\rangle$\tabularnewline
 &  &  & $|1\rangle$  & $|1\rangle$  & $\nicefrac{1}{32}$  & $-$  & $|-\rangle$  & $-$  & $|-\rangle$\tabularnewline
\hline 
 & $|1\rangle$  & $|-\rangle$  & $|1\rangle$  & $|-\rangle$  & $\nicefrac{1}{8}$  & 0  & $|1\rangle$  & 1  & $|1\rangle$\tabularnewline
\cline{2-10}
 &  &  & $|1\rangle$  & $|+\rangle$  & $\nicefrac{1}{64}$  & $-$  & $-$  & 0  & $|+\rangle$\tabularnewline
 & $|+\rangle$  & $|0\rangle$  & $|1\rangle$  & $|-\rangle$  & $\nicefrac{1}{64}$  & 1  & $-$  & 0  & $|+\rangle$\tabularnewline
$|1\rangle$  &  &  & $|0\rangle$  & $|0\rangle$  & $\nicefrac{1}{32}$  & $-$  & $|+\rangle$  & $-$  & $|+\rangle$\tabularnewline
\cline{2-10}
 &  &  & $|1\rangle$  & $|+\rangle$  & $\nicefrac{1}{64}$  & $-$  & $-$  & 1  & $|-\rangle$\tabularnewline
 & $|-\rangle$  & $|1\rangle$  & $|1\rangle$  & $|-\rangle$  & $\nicefrac{1}{64}$  & 1  & $-$  & 1  & $|1\rangle$\tabularnewline
 &  &  & $|0\rangle$  & $|1\rangle$  & $\nicefrac{1}{32}$  & $-$  & $|-\rangle$  & $-$  & $|-\rangle$\tabularnewline
\hline 
 & $|+\rangle$  & $|0\rangle$  & $|+\rangle$  & $|0\rangle$  & $\nicefrac{1}{8}$  & 1  & $|+\rangle$  & 0  & $|+\rangle$\tabularnewline
\cline{2-10}
 &  &  & $|+\rangle$  & $|0\rangle$  & $\nicefrac{1}{64}$  & 0  & $-$  & 0  & $|+\rangle$\tabularnewline
 & $|0\rangle$  & $|+\rangle$  & $|+\rangle$  & $|1\rangle$  & $\nicefrac{1}{64}$  & $-$  & $-$  & 0  & $|0\rangle$\tabularnewline
$|+\rangle$  &  &  & $|-\rangle$  & $|+\rangle$  & $\nicefrac{1}{32}$  & $-$  & $|0\rangle$  & $-$  & $|0\rangle$\tabularnewline
\cline{2-10}
 &  &  & $|+\rangle$  & $|0\rangle$  & $\nicefrac{1}{64}$  & 0  & $-$  & 1  & $|1\rangle$\tabularnewline
 & $|1\rangle$  & $|-\rangle$  & $|+\rangle$  & $|1\rangle$  & $\nicefrac{1}{64}$  & $-$  & $-$  & 1  & $|1\rangle$\tabularnewline
 &  &  & $|-\rangle$  & $|-\rangle$  & $\nicefrac{1}{32}$  & $-$  & $|1\rangle$  & $-$  & $|1\rangle$\tabularnewline
\hline 
 & $|-\rangle$  & $|1\rangle$  & $|-\rangle$  & $|1\rangle$  & $\nicefrac{1}{8}$  & 1  & $|-\rangle$  & 1  & $|-\rangle$\tabularnewline
\cline{2-10}
 &  &  & $|-\rangle$  & $|0\rangle$  & $\nicefrac{1}{64}$  & $-$  & $-$  & 0  & $|0\rangle$\tabularnewline
 & $|0\rangle$  & $|+\rangle$  & $|-\rangle$  & $|1\rangle$  & $\nicefrac{1}{64}$  & 0  & $-$  & 0  & $|0\rangle$\tabularnewline
$|-\rangle$  &  &  & $|+\rangle$  & $|+\rangle$  & $\nicefrac{1}{32}$  & $-$  & $|0\rangle$  & $-$  & $|0\rangle$\tabularnewline
\cline{2-10}
 &  &  & $|-\rangle$  & $|0\rangle$  & $\nicefrac{1}{64}$  & $-$  & $-$  & 1  & $|1\rangle$\tabularnewline
 & $|1\rangle$  & $|-\rangle$  & $|-\rangle$  & $|1\rangle$  & $\nicefrac{1}{64}$  & 0  & $-$  & 1  & $|-\rangle$\tabularnewline
 &  &  & $|+\rangle$  & $|-\rangle$  & $\nicefrac{1}{32}$  & $-$  & $|1\rangle$  & $-$  & $|1\rangle$\tabularnewline
\hline
\end{tabular}
\end{table}

\subsection*{Protocol 1}

To describe these protocols, we utilize the elements of the bases
$Z$ and $X$, along with a notation that defines the basis elements
as $|+z\rangle/|-z\rangle(|+x\rangle/|-x\rangle):=|0\rangle/|1\rangle(|+\rangle/|-\rangle).$
Here, we define the elements of the $Z$ and $X$ bases as

\begin{equation}
\begin{array}{lcl}
|+x\rangle=\frac{1}{\sqrt{2}}\left(|0\rangle+|1\rangle\right) & , & |-x\rangle=\frac{1}{\sqrt{2}}\left(|0\rangle-|1\rangle\right)\\
|+z\rangle=\frac{1}{\sqrt{2}}\left(|+x\rangle+|-x\rangle\right) & , & |-z\rangle=\frac{1}{\sqrt{2}}\left(|+x\rangle-|-x\rangle\right).
\end{array}\label{eq:define=000020the=000020bases=000020elements}
\end{equation}

\begin{description}
\item [{Step~1}] Alice randomly prepares a single qubit sequence $S_{A}$
using $Z$ or $X$ basis and sends it to Bob while keeping the basis
information secret. 
\item [{Step~2}] Bob measures the qubits of sequence $S_{A}$ randomly
in the $Z$ or $X$ basis and records the measurement result. Bob
then prepares a new qubit sequence $S_{B1}$ with the same states
corresponding to the measurement result of the sequence $S_{A}$ and
sends it to Alice. 
\item [{Step~3}] Alice measures each qubit of sequence $S_{B1}$ using
the same basis that was used to prepare the qubit of the sequence
$S_{A}$. For instance, if Alice chooses to prepare the $i^{th}$
qubit of sequence $S_{A}$ in the $Z$ basis ($X$ basis), then she
would measure the $i^{th}$ qubit of sequence $S_{B1}$ using the
$Z$ basis ($X$ basis). Alice records the measurement outcome of
the sequence $S_{B1}$ and asks Bob to proceed if the measurement
outcomes are within the threshold limit of the expected probability
distribution of the possible results. 
\item [{Step~4}] Bob prepares a second qubit sequence $S_{B2}$ with the
same bit values as $S_{B1}$, but using the complementary basis, and
sends the sequence to Alice. For example, if the $i^{th}$ qubit of
$S_{B1}$ is in the state $|\pm z\rangle(|\pm x\rangle),$ then the
$i^{th}$ qubit of $S_{B2}$ will be prepared in the state $\text{\ensuremath{|\pm x\rangle(|\pm z\rangle)}}$
by Bob. 
\item [{Step~5}] Alice performs a measurement on each qubit of $S_{B2}$
based on the measurement result for the qubits of $S_{B1}$. She uses
$X$ basis or $Z$ basis ($Z$ basis or $X$ basis) if she gets the
same state $|\pm z\rangle$ or $|\pm x\rangle$ (states orthogonal
to the initial state, i.e., $|\mp z\rangle$or $|\mp x\rangle$) as
a measurement result of $S_{B1}$ for the corresponding elements to
her initial sequence $S_{A}$. 
\item [{Step~6}] Alice isolates the conclusive measurement results (those
which can be used to definitively determine Bob's measurement results)
obtained from her measurements on the sequence $S_{B1}$ and $S_{B2}$.
If Alice prepares the $i^{th}$ qubit of sequence $S_{A}$ in $|\pm z\rangle(|\pm x\rangle)$
and obtains the measurement result for the corresponding element of
sequence $S_{B1}$ and $S_{B2}$ as $|\mp z\rangle(|\mp x\rangle)$
and $|\pm z\rangle(|\pm x\rangle)$ or $|\mp z\rangle(|\mp x\rangle)$,
respectively, she determines the Bob's measurement result of sequence
$S_{A}$ as $|\pm x\rangle(|\pm z\rangle)$ or $|\mp x\rangle(|\mp z\rangle)$
(see Table \ref{tab:For-protocol-1}).\\
 It may be observed that these conclusive measurements lead to generating
a sifted key without announcing the value of $J$. Step 6 corresponds
to the point (a) mentioned in \emph{Condition for key-sifting.}
\item [{Step~7}] Alice retains those bits as the sifted key for which
the $J$ value is the same for both of parties. For example, if Alice
prepares the sequence $S_{A}$ in the state $|\pm z\rangle(|\pm x\rangle)$
and the measurement result for the corresponding qubit in $S_{B1}$
and $S_{B2}$ are $|\pm z\rangle(|\pm x\rangle)$ and $|\pm x\rangle(|\pm z\rangle)$,
respectively, then Alice determines Bob's measurement result of $S_{A}$
as $|\pm z\rangle(|\pm x\rangle)$ only when the basis used by both
Alice and Bob for preparation and measurement of each element $S_{A}$
is the same (i.e., $J$ value is the same). \\
 It may be noted that a classical sifting process is performed in
this step, corresponds to point (b) in the \emph{Condition for key-sifting}.
This step also contributes to generating the sifted key using the
$J$ value (see Table \ref{tab:For-protocol-1}). The $J$ value is
disclosed only for qubits where Alice's measurement on $S_{B1}$ matches
the corresponding elements of $S_{A}$, provided that the corresponding
qubits in $S_{B2}$ align in same bit elements under a different basis
of $S_{B1}$. 
\end{description}
\begin{table}[H]
\caption{\protect\label{tab:For-protocol-1}Table for mapping between measurement
result and determined result by Alice for Protocol 1}

\centering{}%
\begin{tabular*}{16.5cm}{@{\extracolsep{\fill}}|@{\extracolsep{\fill}}|>{\centering}p{2.5cm}|>{\centering}p{4.5cm}|>{\centering}p{4.5cm}|>{\centering}p{4.25cm}|}
\hline 
$S_{A}$  & Measurement result of $S_{B1}$, $S_{B2}$ by Alice  & Result determined without $J$ value  & Result determined with same $J$ value\tabularnewline
\hline 
$|\pm z\rangle$  & $|\pm z\rangle$,$|\pm x\rangle$  & $-$  & $|\pm z\rangle$\tabularnewline
 & $|\mp z\rangle,$$|\pm z\rangle$ or $|\mp z\rangle,$$|\pm z\rangle$  & $|\pm x\rangle$ or $|\mp x\rangle$ & $-$\tabularnewline
\hline 
$|\pm x\rangle$  & $|\pm x\rangle$,$|\pm z\rangle$  & $-$  & $|\pm x\rangle$\tabularnewline
 & $|\mp x\rangle,$$|\pm x\rangle$ or $|\mp x\rangle,$$|\mp x\rangle$  & $|\pm z\rangle$ or $|\mp z\rangle$ & $-$\tabularnewline
\hline 
\end{tabular*}
\end{table}

\subsection*{Protocol 2}

We introduce a new variable $M\in\{0,1\}$, which will be useful for
interpreting Bob's measurement results of the sequence $S_{A}$. Specifically,
we define: $M(=0):=\{|+z\rangle,|+x\rangle\}$ and $M(=1):=\{|-z\rangle,|-x\rangle\}$
for the classical key-sifting process. Steps 1 to 6 remain the same
for this second protocol, with some differences in the classical sub-protocol
as explained in Step 7. Using this classical sifting process, we obtain
the sifted key with a maximum inherent error having the probability
of $\nicefrac{1}{16}$, but with better efficiency compared to Protocol
1. This trade-off part will be explained later with a detailed analysis. 
\begin{description}
\item [{Step~7}] If Alice prepares the elements of the sequence $S_{A}$
in $|+z\rangle/|+x\rangle(|-z\rangle/|-x\rangle)$ under the following
conditions: $(1)$ Bob announces the value of $M$ as $1(0)$, Alice
determines Bob's measurement result of the sequence $S_{A}$ as $|-x\rangle/|-z\rangle(|+x\rangle/|+z\rangle)$,
irrespective of the measurement result of the sequences $S_{B1}$
and $S_{B2}$, $(2)$ Bob announces the value of $M$ as $0(1)$,
Alice determines Bob's measurement result of the sequence $S_{A}$
as $(i)$ $|+x\rangle/|+z\rangle(|-x\rangle/|-z\rangle)$
if the measurement result of the sequence $S_{B1}$ and $S_{B2}$
are $|+z\rangle/|+x\rangle(|-z\rangle/|-x\rangle)$ and $|-x\rangle/|-z\rangle(|+x\rangle/|+z\rangle)$
respectively, and $(ii)$ $|+z\rangle/|+x\rangle(|-z\rangle/|-x\rangle)$
if the measurement result of the sequence $S_{B1}$ and $S_{B2}$
are $|+z\rangle/|+x\rangle(|-z\rangle/|-x\rangle)$ and $|+x\rangle/|+z\rangle(|-x\rangle/|-z\rangle)$
respectively (see Table \ref{tab:For-protocol-2}). The $M$ value
is revealed only for a subset of qubits where Alice's measurement
on $S_{B1}$ matches the corresponding elements of $S_{A}$, and the
qubits in $S_{B1}$ are in a different basis than their corresponding
elements in $S_{B2}$. After Bob announces the $M$ values of certain
qubits, the encryption criteria shift from the state orientation of
the BB84 protocol to the SARG04 protocol. Specifically, in Protocol
2, the encryption rules are as follows: $|\pm z\rangle$ states encode
the bit value $0$, and $|\pm x\rangle$ states encode the bit value
$1$. As a result, revealing the $M$ values does not provide sufficient
information to accurately infer the final key.
\end{description}
If we consider an inherent error probability of $\frac{1}{16}$, Protocol
2 would yield a higher key rate compared to Protocol 1 in the absence
of Eve. In this context, Protocol 2 and Protocol 1 demonstrate efficiencies
of 0.192 and 0.2069, respectively. Notably, unlike Protocol 2, Protocol
1 does not introduce inherent errors. A detailed analysis of these
results, along with a discussion of the associated trade-offs, is
provided in Appendix F. 
\begin{table}
\caption{\protect\label{tab:For-protocol-2}Table for mapping between measurement
result and determined result by Alice for protocol 2}

\centering{}%
\begin{tabular*}{16.5cm}{@{\extracolsep{\fill}}|@{\extracolsep{\fill}}|>{\centering}p{2.75cm}|>{\centering}p{2.5cm}|>{\centering}p{3.5cm}|>{\centering}p{3.4cm}|>{\centering}p{3.4cm}|}
\hline 
$S_{A}$ & Value of $M$ & Measurement result of $S_{B1}$ by Alice & Measurement result of $S_{B2}$ by Alice & Result determined\tabularnewline
\hline 
 & 1 & $-$ & $-$ & $|-x\rangle/|-z\rangle$\tabularnewline
\cline{2-5}
$|+z\rangle/|+x\rangle$ & 0 & $|+z\rangle/|+x\rangle$ & $|-x\rangle/|-z\rangle$ & $|+x\rangle/|+z\rangle$\tabularnewline
 & 0 & $|+z\rangle/|+x\rangle$ & $|+x\rangle/|+z\rangle$ & $|+z\rangle/|+x\rangle$\tabularnewline
\hline 
 & 0 & $-$ & $-$ & $|+x\rangle/|+z\rangle$\tabularnewline
\cline{2-5}
$|-z\rangle/|-x\rangle$ & 1 & $|-z\rangle/|-x\rangle$ & $|+x\rangle/|+z\rangle$ & $|-x\rangle/|-z\rangle$\tabularnewline
 & 1 & $|-z\rangle/|-x\rangle$ & $|-x\rangle/|-z\rangle$ & $|-z\rangle/|-x\rangle$\tabularnewline
\hline 
\end{tabular*}
\end{table}

\section{Security performance for the proposed protocols\protect\label{sec:III}}

We previously mentioned that Alice's approval of the sequence $S_{B1}$
is a prerequisite before Bob can proceed to transmit the sequence
$S_{B2}$. Upon receiving Alice's acceptance of $S_{B1}$, Bob proceeds
to transmit the second sequence $S_{B2}$. Ultimately, Alice and Bob
reach a consensus on the secret key, provided that the calculated
error percentage falls below the acceptable error threshold following
the successful completion of the protocol. The primary objective of
our security analysis for the proposed protocols is to determine the
maximum allowable error under the presence of a series of collective
attacks. To understand Eve's potential attack strategy, we employ
a methodology inspired by the approach outlined in Ref. \citep{KGR_05},
which involves the use of a depolarizing map capable of transforming
any two-qubit state into a Bell-diagonal state. If we intend to evaluate
the security of the QKD protocols introduced here in alignment with
the principles presented in Ref. \citep{KGR_05}, we must adapt our
protocols to equivalent entanglement-based schemes. A corresponding
approach to Protocol 1/2, as described earlier, can be visualized
as follows: Alice generates a set of $n$ two-qubit entangled states
(for instance, Bell states) and applies her encoding procedure to
the first qubit of each pair, while sending the second qubit to Bob.
In other words, if Alice prepares a state like $|\Phi^{+}\rangle$,
she modifies it into $A_{j}\otimes I_{2}|\Phi^{+}\rangle$ and forwards
the second qubit to Bob. Here, $|\Phi^{\pm}\rangle$ signifies $\frac{1}{\sqrt{2}}(|00\rangle\pm|11\rangle)$,
and the operators $A_{j}$ and $I_{2}$ represent Alice's encoding
operation and the identity operation in a two-dimensional space, respectively.
Bob also randomly applies one of his encoding operators $B_{j}$ to
each of the qubits that he receives. We can denote the $2n$ qubit
state shared by Alice and Bob as $\tilde{\rho}_{AB}^{n}$. Finally,
Alice and Bob measure their qubits of $\tilde{\rho}_{AB}^{n}$ randomly
in $X$ and $Z$ bases and map each measurement outcome to bit value
0 or 1. We use two completely positive maps (CPMs), $\mathcal{O}_{1}$
and $\mathcal{O}_{2}$, where $\mathcal{O}_{1}$ is entirely defined
by the protocol, and $\mathcal{O}_{2}$ is independent of the protocol.
Specifically, these CPMs are defined as $\mathcal{O}_{1}(\rho)=\frac{1}{N}\sum_{j}p_{j}A_{j}\otimes B_{j}(\rho)A_{j}^{\dagger}\otimes B_{j}^{\dagger}$
and $\mathcal{O}_{2}(\rho)=\sum_{l}M_{l}\otimes M_{l}(\rho)M_{l}^{\dagger}\otimes M_{l}^{\dagger}$.
Here, $p_{j}\ge0$ is the probability that Alice and Bob decide to
keep the bit value during the sifting subprotocol, $N$ is the normalization
factor, and $M_{l}$ describes a quantum operation such that $M_{l}\in\left\{ I_{2},\sigma_{x},\sigma_{y},\sigma_{z}:I_{2}=|0\rangle\langle0|+|1\rangle\langle1|,\sigma_{x}=|0\rangle\langle1|+|1\rangle\langle0|,i\sigma_{y}=|0\rangle\langle1|+|1\rangle\langle0|,\sigma_{z}=|0\rangle\langle0|-|1\rangle\langle1|\right\} $.
The structure of $\mathcal{O}_{2}(\rho)$ shows that the same operator
is applied on both qubits, thus $M_{l}\otimes M_{l}\in\left\{ I\otimes I,\,\sigma_{x}\otimes\sigma_{x},\,\sigma_{y}\otimes\sigma_{y},\,\sigma_{z}\otimes\sigma_{z}\right\} $.
These two-qubit operators are applied with equal probability, or equivalently,
these are applied randomly. Interestingly, the random application
of these operations mimics the action of a depolarizing channel that
transforms any two-qubit state to a Bell diagonal state. If Alice
and Bob apply unitary operation $A_{j}\otimes B_{j},$\footnote{We may define the encoding and decoding operation in generalized form
as $A_{j}=|0\rangle\langle(\phi_{j}^{0})^{*}|+|1\rangle\langle(\phi_{j}^{1})^{*}|$
and $B_{j}=|0\rangle\langle(\phi_{j}^{1})^{\perp}|+|1\rangle\langle(\phi_{j}^{0})^{\perp}|$,
where $|(\phi_{j}^{i})^{*}\rangle$ denotes the complex conjugate
state of $|\phi_{j}^{i}\rangle$ and $|(\phi_{j}^{i})^{\perp}\rangle$
denotes the orthogonal state to $|\phi_{j}^{i}\rangle$ in computational
basis, $j\in\{1,\cdots,m\}$ is the set of states used to encode the
bit values $i=0,1$ \cite{RGK_05,KGR_05}.} they get their sifted key after the sifting phase with the normalization
factor $N$, where $\sum_{j}p_{j}=1$. We use a normalized two-qubit
density operator from Eq (1) of Ref. \citep{RGK_05} with $n=1$ (see
for details \citep{KGR_05}). We use the notation $P_{|\Phi\rangle}=|\Phi\rangle\langle\Phi|$
which describes a state projection operator that projects a quantum
state of the same dimension onto the state $|\Phi\rangle$. Here,
\begin{equation}
\rho^{1}[\boldsymbol{\mathcal{\mu}}]=\mu_{1}P_{|\Phi^{+}\rangle}+\mu_{2}P_{|\Phi^{-}\rangle}+\mu_{3}P_{|\Psi^{+}\rangle}+\mu_{4}P_{|\Psi^{-}\rangle},\label{eq:1}
\end{equation}
where $P_{|\Phi^{\pm}\rangle}$ and $P_{|\Psi^{\pm}\rangle}$ are
the state projection operators onto the Bell states $|\Phi^{\pm}\rangle\text{=\ensuremath{\frac{1}{\sqrt{2}}(|00\rangle\pm|11\rangle})}$
and $|\Psi^{\pm}\rangle$$=\ensuremath{\frac{1}{\sqrt{2}}(|01\rangle\pm|10\rangle})$
and $\mu_{1/2}$ and $\mu_{3/4}$ are the respective probabilities
of obtaining the corresponding Bell states in the depolarizing channel.
In what follows, to analyze the security of the sequence $S_{B1}$,
we use the following key rate equation for one-way quantum channel
\begin{equation}
r:=I(A:B)-\underset{\rho\in\mathscr{\mathcal{R}}}{max}S(\rho),\label{eq:2.}
\end{equation}
where $A$ and $B$ are the quantum states obtained after the measurements
are performed by Alice and Bob, $I(A:B)$ is the mutual information
between Alice and Bob, $S(\rho)$ is the von Neumann entropy of the
composite state of both the parties (i.e., $\rho$) and $\mathcal{R}$
is the density range\footnote{Let $\mathcal{S}(\mathcal{H})$ denote the set of density operators
on $\mathcal{H}\equiv H_{\textrm{A}}\otimes\mathcal{H}_{\textrm{B}}$.
Consider a density operator $\rho^{\prime}$ on $\mathcal{H}^{\otimes n}$,
i.e., $\rho^{\prime}\in\mathcal{S}(\mathcal{H}^{\otimes n})$, with
a density range $\mathcal{R}\subseteq\mathcal{S}(\mathcal{H})$. The
density range, $\mathcal{R}$, represents the set of reduced density
operators on individual subsystems derived from $\rho^{\prime}$.
Specifically,$\mathcal{R}:=\mathcal{R}(a,b)$ is defined as the set
of density operators on $\mathcal{H_{\textrm{A}}\otimes\mathcal{H}_{\textrm{B}}}$
such that the measurement outcomes for any $\rho\in\mathcal{R}$ correspond
to Alice's and Bob's measurement operators \cite{CRE_04}.} of the density operator $\rho$ (for a precise definition of density
range, see Definition 3.16 of \cite{CRE_04}). This equation was introduced
in Ref. \cite{CRE_04} (see Eq. (22) of \cite{CRE_04}). Eq. (\ref{eq:2.})
does not directly provide the key rate for our protocols; rather,
it is used to determine the secure error limit for $S_{B1}$. Since
$S_{B1}$ is transmitted via a one-way quantum channel (from Bob to
Alice), this key rate equation is applied. Bob partially announces
information about his measurement results for the sequence $S_{A}$
(or the prepared states of $S_{B1}$), allowing both parties to establish
correlations for the security check of $S_{B1}$ after this classical
announcement. When Alice knows the elements of her initial sequence
$S_{A}$, she can use Eq. (\ref{eq:2.}) to calculate the tolerable
error limit for $S_{B1}$. Let us now assume that the quantum bit
error rate (QBER) is $\mathcal{E}\in[0,1]$ for the measurements done
in both $X$ and $Z$ bases. The outcome for the projective measurements
on the system $\rho$ can be captured through a random variable $V$.
As the measurement in the bases $Z$ and $X$ can lead to four different
outcomes, we can have four probabilities associated with these measurement
outcomes. In fact, the probabilities of the measurement outcome in
the bases $Z$ and $X$ can be defined as the probabilities ($\mu_{i}:i\in\{1,2,3,4\}$)
of obtaining different values of $V$. The entropy of this variable
$V$ is $H(V)=-\sum_{V\in\mu_{i}}V\log_{2}V\ge S(\rho)$. These probabilities
$\mu_{i}{\rm s}$ can be computed easily by taking expectation values
of $\rho$ with respect to the relevant states. For example, in our
case, $\mu_{1}=\langle\Phi^{+}|\rho|\Phi^{+}\rangle,\,\mu_{2}=\langle\Phi^{-}|\rho|\Phi^{-}\rangle,\,\mu_{3}=\langle|\Psi^{+}|\rho|\Psi^{+}\rangle$
and $\mu_{4}=\langle|\Psi^{-}|\rho|\Psi^{-}\rangle.$ Through a long
but straightforward calculation, we obtain relations between the probabilities
$\mu_{i}$s associated with the system described in Eq. (\ref{eq:1})
as follows: $\mu_{3}+\mu_{4}=\mathcal{E}$, $\mu_{2}+\mu_{4}=\mathcal{E}$,
$\mu_{1}+\mu_{2}=1-\mathcal{E}$ and $\mu_{1}+\mu_{3}=1-\mathcal{E}$.
These four equations are not linearly independent. There are actually
three linearly independent equations (for example, you may consider
(i) the first three of these equations or (ii) the first two and the
last one as linearly independent equations). In this situation, we
cannot solve the above set of equations directly, but we can consider
one of the probabilities as a free parameter and express the rest
of the probabilities in terms of it. Here we choose $\mu_{4}$ as
the free parameter to express other probabilities in terms of it as
$\mu_{1}=1-2\mathcal{E}+\mu_{4}$ and $\mu_{2}=\mu_{3}=\mathcal{E}-\mu_{4}$.
It may be noted that $\mu_{4}\in[0,\mathcal{E}]$ as the range of
any probability is $[0,1]$ and $\mu_{2}+\mu_{4}=\mathcal{E}$ (refer
to Appendix A for more information). It is important to note that
the proposed schemes prioritize a larger quantum component and a smaller
classical component. Our analysis focuses on investigating the information-theoretic
security bounds of each quantum sequence. If the QBER is within the tolerable secure limit, both parties proceed
with the subsequent steps of the protocol. The following analysis
primarily applies to both new protocols, as all steps prior to the
classical sub-protocol (the quantum part) are identical for both.
For simplicity, the security analysis is limited to the quantum component
of the proposed scheme.

\begin{figure}
\begin{centering}
\includegraphics[scale=0.5]{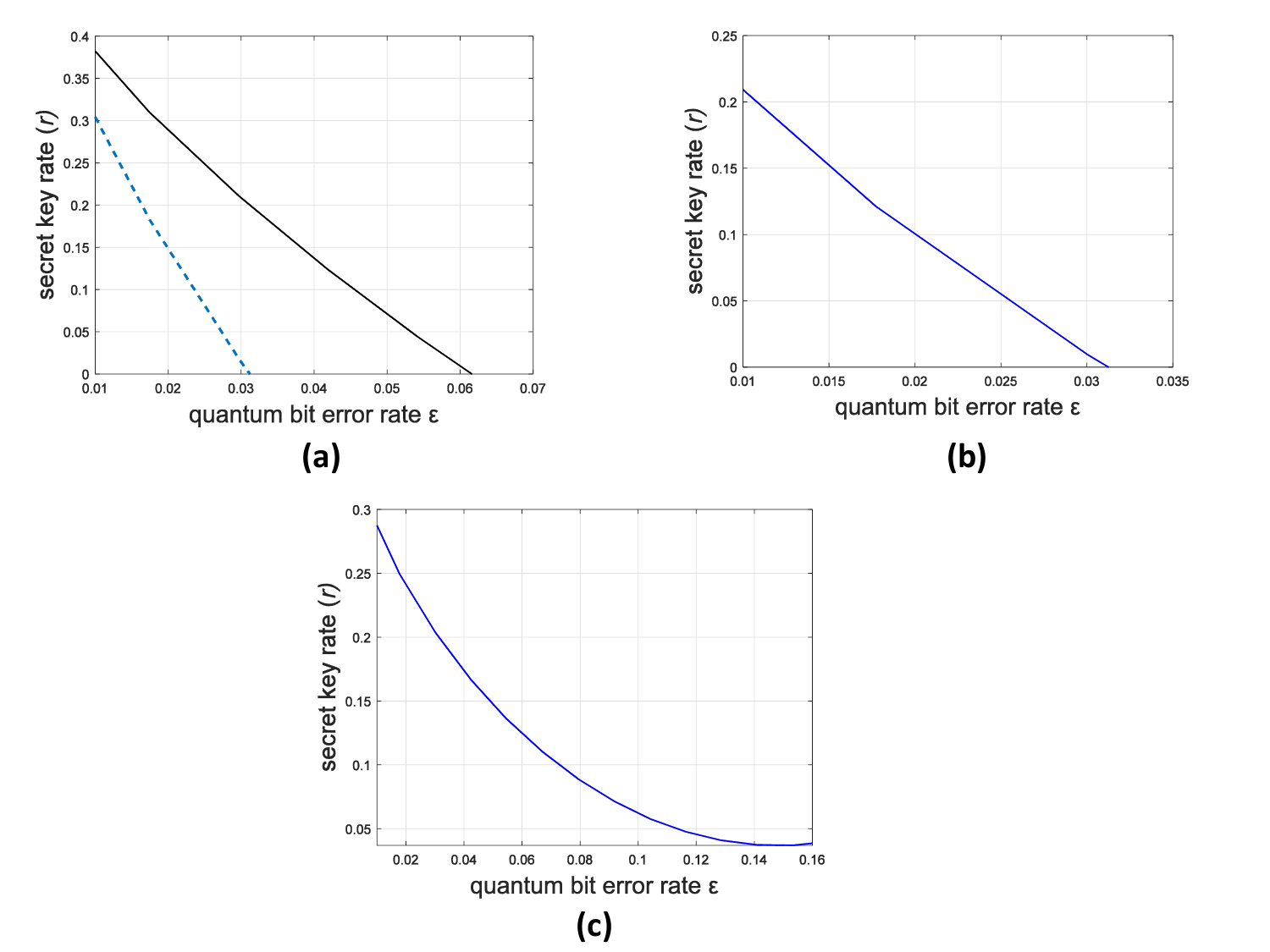} 
\par\end{centering}
\caption{\protect\label{fig:maximum=000020tolerable=000020error=000020limit}(Color
online) Plot of the secret key rate as a function of quantum bit error
rate $\mathcal{E}$: $(a)$ solid (black) line and dashed (blue) line
illustrate the maximum tolerable error limit (security threshold)
evaluation for the sequence $S_{B1}$ with and without the introduction
of the new variable $\mathcal{Y}$, respectively, $(b)$ plot evaluates
the maximum tolerable error limit (security threshold) for the sequence
$S_{B2}$ without introducing the new variable $\mathcal{X}$, and
$(c)$ plot evaluates the maximum tolerable error limit (security
threshold) for sequence $S_{B2}$ with the introduction of the new
variable $\mathcal{X}$.}
\end{figure}

The following analysis determines the secure error limit for the sequence
$S_{B1}$ with and without classical pre-processing \cite{CRE_04}.
To maximize the entropy of the random variable $V$, solve $\frac{d\left(H\left(V\right)\right)}{d\mu_{4}}=0$.
This yields $\mu_{4}=\mathcal{E}^{2}$, and the corresponding entropy
$H(V)$ becomes $2h(\mathcal{E})$, where $h(\mathcal{E})=-\mathcal{E}\log_{2}\mathcal{E}-(1-\mathcal{E})\log_{2}(1-\mathcal{E})$
is the binary entropy function. The entropy of Bob's measurement results
is $H(B)=2$, and the conditional entropy of $B$ given $A$ is $H(B|A)=1-\frac{1-\mathcal{E}}{2}\log_{2}\frac{1-\mathcal{E}}{2}$$-\frac{\mathcal{E}}{2}\log_{2}\frac{\mathcal{E}}{2}$.
The security threshold (or maximum tolerable error limit) is defined
as the largest value of $\mathcal{E}$ for which Eq. (\ref{eq:2.})
remains positive for $S_{B1}$. Under these conditions, solving: $1+\frac{1-\mathcal{E}}{2}\log_{2}\frac{1-\mathcal{E}}{2}+\frac{\mathcal{E}}{2}\log_{2}\frac{\mathcal{E}}{2}-2h(\mathcal{E})=0$
yields $\mathcal{E}\approx0.0314$ (i.e., $3.14\%$ QBER; see Appendix
A and Appendix B for details). This implies that, without classical
pre-processing, the maximum tolerable theoretical error limit for
$S_{B1}$ is 3.14\% after Bob's classical announcement and the evaluation
of correlations for the security check. To improve the security threshold,
we introduce a new variable $\mathscr{\mathcal{Y}}=j_{A}\oplus j_{B}$\footnote{It can be viewed as a classical pre-processing method that helps to
increase the key-rate as well as maximum tolerable error limit \cite{CRE_04}.
For the same purpose $\mathcal{X}$ is also used in context of the
sequence $S_{B2}$.}, where $j_{A}$ and $j_{B}$ are the bases chosen by Alice and Bob
to measure the particles in $S_{B1}$ and $S_{A}$, respectively,
with $j_{A},j_{B}\in\{0,1\}$. Solving\footnote{One needs to replace $H(V)$ with $H(V)-h(\mathcal{E})$ (see Appendix
D) in the results presented in Appendix B.}: $1+\frac{1-\mathcal{E}}{2}\log_{2}\frac{1-\mathcal{E}}{2}+\frac{\mathcal{E}}{2}\log_{2}\frac{\mathcal{E}}{2}-h(\mathcal{E})=0$
gives $\mathcal{E}\approx0.0617$ (i.e., $6.17\%$ bit error rate;
see Fig. \ref{fig:maximum=000020tolerable=000020error=000020limit}
(a)). This result shows that, with $\mathcal{Y}$ announced, the maximum
tolerable error limit for measuring $S_{A}$ and $S_{B1}$ increases
to $6.17\%$. Thus, with classical pre-processing, the maximum tolerable
theoretical error limit for $S_{B1}$ is $6.17\%$. In Section \ref{sec:II},
we noted that the probabilities of obtaining expected outcomes from
$S_{A}$ are $\frac{1}{2}$ ($\text{\ensuremath{\psi_{Z}^{x}}}$)
for the same basis and $\frac{1}{4}$ ($\psi_{X}^{x/x^{\perp}}$)
for different basis. For $S_{B1}$, the probability is $\frac{1}{4}$
($\psi_{J}^{x^{\perp}}$). The maximum tolerable error percentages
for deviations from these probabilities are $3.14\%$ without announcing
$\mathcal{Y}$ and $6.17\%$ with $\mathcal{Y}$ announced. The introduction
of the new variables $\mathcal{Y}$ (and $\mathcal{X}$, introduced
in the next paragraph) is inspired by the proposal of Christandl et
al. \cite{CRE_04}. The method for determining these variables is
based on the principles of information reconciliation and privacy
amplification against quantum adversaries. In information reconciliation,
hash functions and guessing functions are employed to derive these
variables. For privacy amplification, the process involves a hash
function and a function of the measurement outcomes of quantum states
relative to an arbitrary POVM, which also depends on the hash function
(see 4.2, 4.3 and 5.1 in \cite{CRE_04}).

Alice starts by checking the security threshold for the sequence $S_{B1}$,
ensuring it falls within the expected limit. Subsequently, she proceeds
to measure the second sequence, $S_{B2}$, transmitted by Bob, completing
the sifting subprotocol. Following this key sifting process, Alice
and Bob then evaluate whether the QBER is below the security threshold.
The determination of the acceptable threshold value follows the same
procedure as previously described. The entropy of Bob's final bit
string $b$ after the sifting subprotocol is denoted as $H(b)=1$,
and the conditional entropy of bit string $b$ when Alice's bit string
$a$ is known is calculated as $H(b|a)=h(\frac{1}{6}+\frac{2\mathcal{E}}{3})$,
where $a,b\in\{0,1\}$. In a similar fashion, we obtain the equation
for the positive key rate for one-way quantum channel to get tolerable
QBER of $S_{B2}$. 
\begin{equation}
1-h(\frac{1}{6}+\frac{2\mathcal{E}}{3})-2h(\mathcal{E})=0,\label{eq:new=0000201}
\end{equation}
(i.e., by considering $r=0)$ and solving it, we can obtain the security
threshold as $\mathcal{E}\approx0.0316$ (see Fig. \ref{fig:maximum=000020tolerable=000020error=000020limit}
(b)) i.e., $3.16\%$ QBER (see Appendix C for more information). To
improve the security threshold, one can introduce a random variable
$\mathcal{X}=a\oplus b$ that contains information about the error
position. The introduction of $\mathcal{X}$ decreases the quantum
part (last part) of the Eq. (\ref{eq:2.}) but not the minimum entropy
value of string $b$ (for details see Sec. 5.1 of Ref. \cite{CRE_04}).
To elaborate on this point, we can divide the quantum system into
four subsystems, each corresponding to an error and no-error situation
for each basis. For basis $Z(X)$, the error and no-error comprise
fractions of $\frac{\mathcal{E}}{2}$ and $\frac{1-\mathcal{E}}{2}$
of the total number of qubits, respectively. After calculating the
entropy of the four subsystems in the error and no-error scenarios,
one can obtain $h\left(\frac{\mathcal{E}-\mu_{4}}{\mathcal{E}}\right)$
and $h\left(\frac{1-2\mathcal{E}+\mu_{4}}{1-\mathcal{E}}\right)$
as entropy for error and no-error situations. After performing the
statistical averaging over the four possible subsystems, we obtain
(see Appendix D for a more comprehensive calculation)

\begin{equation}
(1-\mathcal{E})h\left(\frac{1-2\mathcal{E}+\mu_{4}}{1-\mathcal{E}}\right)+\mathcal{E}h\left(\frac{\mathcal{E}-\mu_{4}}{\mathcal{E}}\right)=H(V)-h(\mathcal{E}).\label{eq:reconditioned=000020entropy}
\end{equation}
We can substitute this reconditioned entropy for variable $V$ in
Eq. (\ref{eq:new=0000201}) to obtain a modified key rate equation
for one-way quantum channel to get tolerable QBER of $S_{B2}$: 
\begin{equation}
r=1-h(\frac{1}{6}+\frac{2\mathcal{E}}{3})-h(\mathcal{E}).\label{eq:modified-keyrate}
\end{equation}
here, the solution of the equation for positive $r$ is the security
threshold, $\mathcal{E}\approx0.15$. Thus, the corresponding new
bit error rate would be $15\%$ (refer to Fig. \ref{fig:maximum=000020tolerable=000020error=000020limit}
(c)).

We can now analyze the secret-key rate under the assumption that the
protocol remains secure against collective attacks by Eve. Firstly,
we describe the initial state $\rho_{AB}^{n}$, which depends on the
threshold QBER at which the protocol does not terminate prematurely.
The state $\rho_{AB}^{n}$ represents a quantum state ideally shared
exclusively between Alice and Bob but is partially accessible to Eve,
who can potentially perform collective attacks on it. Let $\Gamma$
be the collection of all two-qubit states $\sigma_{AB}$ that can
result from Eve's collective attack on the initial state $\rho_{AB}^{n}$.
The success of the attack depends on it leaving no discernible traces.
In such a scenario, we must have $\sigma_{AB}^{\otimes n}=\rho_{AB}^{n}$.
However, the attack may not always succeed; in cases where it fails,
it leaves detectable traces, leading to the termination of the protocol.
Our interest lies in situations where the protocol is not terminated.
To account for the possibility of such a situation, we assume the
existence of a protocol (operation) that Eve can utilize to produce
a state $\sigma_{AB}^{\otimes n}=\rho_{AB}^{n}$ using ancillary qubits
and a portion of the initial state shared by Alice and Bob, which
is accessible to Eve through the channel. Following the approach in
Ref. \cite{RGK_05}, we can define a set $\Gamma_{QBER}$ as a subset
of $\Gamma$, containing all states $\sigma_{AB}$ for which the protocol
does not terminate prematurely. In other words, if $\sigma_{AB}\in\Gamma_{QBER}$,
then the protocol is expected to generate a secret key. Renner et
al. in \cite{RGK_05} have demonstrated that, based on the conditions
outlined above, it is possible to establish both a lower bound and
an upper bound on the secret-key rate for any protocol involving one-way
post-processing.

\begin{equation}
r\ge\underset{c\leftarrow a}{sup}\underset{\sigma_{AB}\in\Gamma_{QBER}}{inf}\left(S(c|E)-H(c|b)\right).\label{eq:4}
\end{equation}
here $r_{c\leftarrow a}$ is the rate that can be achieved if the
channel\footnote{$c\leftarrow a$ may be visualized as $q(|1\rangle_{ca}\langle0|+|0\rangle_{ca}\langle1|)+(1-q)(|0\rangle_{ca}\langle0|+|1\rangle_{ca}\langle1|)$,
where $a$ denotes Alice's register of classical outcome and $c$
denotes the register of the noisy version of $a$.} $c\leftarrow a$ is used for the pre-processing, $S(c|E)$ denotes
the von Neumann entropy of $c$ conditioned on Eve's initial state,
i.e., $S(c|E)=S(\sigma_{cE})-S(\sigma_{E})$. This state $\sigma_{cE}$
is obtained from the two-qubit state $\sigma_{AB}$ by taking a purification
$\sigma_{ABE}$ of the Bell diagonal state $\sigma_{AB}^{diag}:=\mathcal{O}_{2}(\sigma_{AB})$.
The state $\sigma_{AB}^{diag}$ has the same diagonal elements as
$\sigma_{AB}$ with respect to the Bell basis. Here, $a$, $b$ and
$e$ are the outcomes of Alice, Bob and Eve's after the measurement
is applied to the first, second and third subsystem of $\sigma_{ABE}$.

To establish the upper limit for the rate, it suffices to focus exclusively
on collective attacks. The composite system involving Alice, Bob and
Eve exhibits a product structure denoted as $\rho_{ABE}^{n}:=\sigma_{ABE}^{\otimes n}$,
where $\sigma_{ABE}$ represents a tripartite state. The $n-$fold
product state $\sigma_{abE}^{n}$ fully characterizes the scenario
in which the single state $\sigma_{abE}$ is obtained when Alice and
Bob perform measurements on the $\sigma_{ABE}$ state (for a detailed
proof, please refer to Section IV of Ref.\cite{RGK_05}). Consequently,
the upper limit on the secret key rate is as follows:

\begin{equation}
r(a,b,e)=\underset{c\leftarrow a}{sup}\left(H(c|e)-H(c|b)\right).\label{eq:5}
\end{equation}
This equation implies that if the supremum is taken over all the channels
(including both quantum and classical channels) $c\leftarrow a$,
it will be the upper bound on the secret key rate.

Now, we analyze our protocol in the context of lower bound and upper
bound of the secret key rate. As before, we take $n=1$, $\sigma_{AB}=\rho^{1}[\mu]$.
It is required to consider a purification $|\Psi\rangle_{ABE}$ of
the Bell diagonal state $\mathcal{O}_{2}(\sigma_{AB})$ originated
from $\sigma_{AB}$ that can be written as follows:

\begin{equation}
|\Psi\rangle_{ABE}:=\stackrel[i=1]{4}{\sum}\sqrt{\mu_{i}}|\varphi_{i}\rangle_{AB}\otimes|\varepsilon_{i}\rangle_{E},\label{eq:Composite_state_with_Eve}
\end{equation}
where $|\varphi_{i}\rangle_{AB}$ denotes the Bell states corresponding
to the joint system of Alice and Bob \footnote{Alice measures her qubit with $Z$ basis, and Bob measures with $Z$
or $X$ basis with $\frac{1}{2}$ probability.}, and $|\varepsilon_{i}\rangle_{E}$ denotes some mutually orthogonal
states in Eve's system, which form the basis $\varepsilon_{E}\in\left\{ |\varepsilon_{1}\rangle_{E},\ldots,|\varepsilon_{4}\rangle_{E}\right\} $.
It can be easily verified that Alice measures her qubit with the $Z(X)$
basis and Bob measures his qubit with the $Z$ or $X$ basis with
equal probability, resulting in the outcomes $|\mathcal{A\rangle}$
and $|\mathcal{B\rangle}$ for Alice and Bob, respectively. For example,
we consider $|\mathcal{A\rangle}\in\left\{ |0\rangle,|1\rangle\right\} $
and $|\mathcal{B\rangle}\in\left\{ |0\rangle,|1\rangle,|+\rangle,|-\rangle\right\} $.
Under this consideration, Eve's state will be $|\phi^{\mathcal{A},\mathcal{B}}\rangle$,
where

\begin{equation}
\begin{array}{lcl}
|\phi^{0,0}\rangle & = & \frac{1}{\sqrt{2}}\left(\sqrt{\mu_{1}}|\varepsilon_{1}\rangle_{E}+\sqrt{\mu_{2}}|\varepsilon_{2}\rangle_{E}\right),\\
|\phi^{1,1}\rangle & = & \frac{1}{\sqrt{2}}\left(\sqrt{\mu_{1}}|\varepsilon_{1}\rangle_{E}-\sqrt{\mu_{2}}|\varepsilon_{2}\rangle_{E}\right),\\
|\phi^{0,1}\rangle & = & \frac{1}{\sqrt{2}}\left(\sqrt{\mu_{3}}|\varepsilon_{3}\rangle_{E}+\sqrt{\mu_{4}}|\varepsilon_{4}\rangle_{E}\right),\\
|\phi^{1,0}\rangle & = & \frac{1}{\sqrt{2}}\left(\sqrt{\mu_{3}}|\varepsilon_{3}\rangle_{E}-\sqrt{\mu_{4}}|\varepsilon_{4}\rangle_{E}\right),\\
|\phi^{0,+}\rangle & = & \frac{1}{2}\left(\sqrt{\mu_{1}}|\varepsilon_{1}\rangle_{E}+\sqrt{\mu_{2}}|\varepsilon_{2}\rangle_{E}+\sqrt{\mu_{3}}|\varepsilon_{3}\rangle_{E}+\sqrt{\mu_{4}}|\varepsilon_{4}\rangle_{E}\right),\\
|\phi^{0,-}\rangle & = & \frac{1}{2}\left(\sqrt{\mu_{1}}|\varepsilon_{1}\rangle_{E}+\sqrt{\mu_{2}}|\varepsilon_{2}\rangle_{E}-\sqrt{\mu_{3}}|\varepsilon_{3}\rangle_{E}-\sqrt{\mu_{4}}|\varepsilon_{4}\rangle_{E}\right),\\
|\phi^{1,+}\rangle & = & \frac{1}{2}\left(\sqrt{\mu_{1}}|\varepsilon_{1}\rangle_{E}-\sqrt{\mu_{2}}|\varepsilon_{2}\rangle_{E}+\sqrt{\mu_{3}}|\varepsilon_{3}\rangle_{E}-\sqrt{\mu_{4}}|\varepsilon_{4}\rangle_{E}\right),\\
|\phi^{1,-}\rangle & = & \frac{1}{2}\left(-\sqrt{\mu_{1}}|\varepsilon_{1}\rangle_{E}+\sqrt{\mu_{2}}|\varepsilon_{2}\rangle_{E}+\sqrt{\mu_{3}}|\varepsilon_{3}\rangle_{E}-\sqrt{\mu_{4}}|\varepsilon_{4}\rangle_{E}\right).
\end{array}
\end{equation}

We are now equipped to compute the density operators of Eve's system
when Alice gets the outcomes $0$ and $1$, denoted as $\sigma_{E}^{0}$
and $\sigma_{E}^{1}$, respectively. Here, we consider the system
accepted by Alice and Bob after classical pre-processing of the protocol,
given by $\sigma_{E}^{0}=\frac{1}{2}\left(P_{|\phi^{0,0}\rangle}+P_{|\phi^{0,1}\rangle}\right)+\frac{1}{2}\left(P_{|\phi^{0,+}\rangle}+P_{|\phi^{0,-}\rangle}\right)$
and $\sigma_{E}^{1}=\frac{1}{2}\left(P_{|\phi^{1,0}\rangle}+P_{|\phi^{1,1}\rangle}\right)+\frac{1}{2}\left(P_{|\phi^{1,+}\rangle}+P_{|\phi^{1,-}\rangle}\right)$
(for a more comprehensive calculation, refer to Appendix E). We can
now obtain the state of Eve with respect to the basis $|\varepsilon_{i}\rangle_{E}$,
where $i\in\{1,\cdots,4\}$ as

\begin{equation}
\sigma_{E}^{k}=\left(\begin{array}{cccc}
\mu_{1} & (-1)^{k}\sqrt{\mu_{1}\mu_{2}} & 0 & 0\\
(-1)^{k}\sqrt{\mu_{1}\mu_{2}} & \mu_{2} & 0 & 0\\
0 & 0 & \mu_{3} & (-1)^{k}\sqrt{\mu_{1}\mu_{2}}\\
0 & 0 & (-1)^{k}\sqrt{\mu_{1}\mu_{2}} & \mu_{4}
\end{array}\right),\label{eq:8}
\end{equation}
where $k\in\left\{ 0,1\}\right\} .$

We have already mentioned channel $c\leftarrow a$ which provides
a noisy version of $a$. We may consider that Alice uses bit-flip
with probability $q$ to make $c$, i.e., $p_{c|a=0}(1)=p_{c|a=1}(0)=q$.
We may now use the following standard relations to simplify the right-hand
side of Eq. (\ref{eq:4}),

\begin{equation}
\begin{array}{lcl}
S(c|E) & = & S(cE)-S(E)\\
 & = & \left[H(c)+S(E|c)-S(E)\right],
\end{array}\label{eq:9}
\end{equation}
and 
\begin{equation}
\begin{array}{lcl}
H(c|b) & = & H(cb)-H(b)\\
 & = & \left[H(c)+H(b|c)-H(b)\right].
\end{array}\label{eq:10}
\end{equation}
Substituting Eq. (\ref{eq:9}) and Eq. (\ref{eq:10}) into the right-hand
side of Eq. (\ref{eq:4}), we can express the entropy difference as
follows:

\begin{equation}
S(c|E)-H(c|b)=S(E|c)-S(E)-\left(H(b|c)-H(b)\right).\label{eq:eleven}
\end{equation}
The above substitution will modify Eq. (\ref{eq:4}) in a manner that
allows us to compute the lower bound of the secret key rate of our
protocol.

If only Eve's system is used for the calculation of the entropy, there
are only two possibilities, where Alice can have $0$ and $1$ bit
value. At the same time, obtaining the entropy of $E$ conditioned
on the value $c$, announced by Alice, depends on the bit-flip probability.
So we have,

\[
S(E|c)=\frac{1}{2}S\left((1-q)\sigma_{E}^{0}+q\sigma_{E}^{1}\right)+\frac{1}{2}S\left(q\sigma_{E}^{0}+(1-q)\sigma_{E}^{1}\right),
\]
and

\[
S(E)=S\left(\frac{1}{2}\sigma_{E}^{0}+\frac{1}{2}\sigma_{E}^{1}\right).
\]

\begin{figure}
\begin{centering}
\includegraphics[scale=0.5]{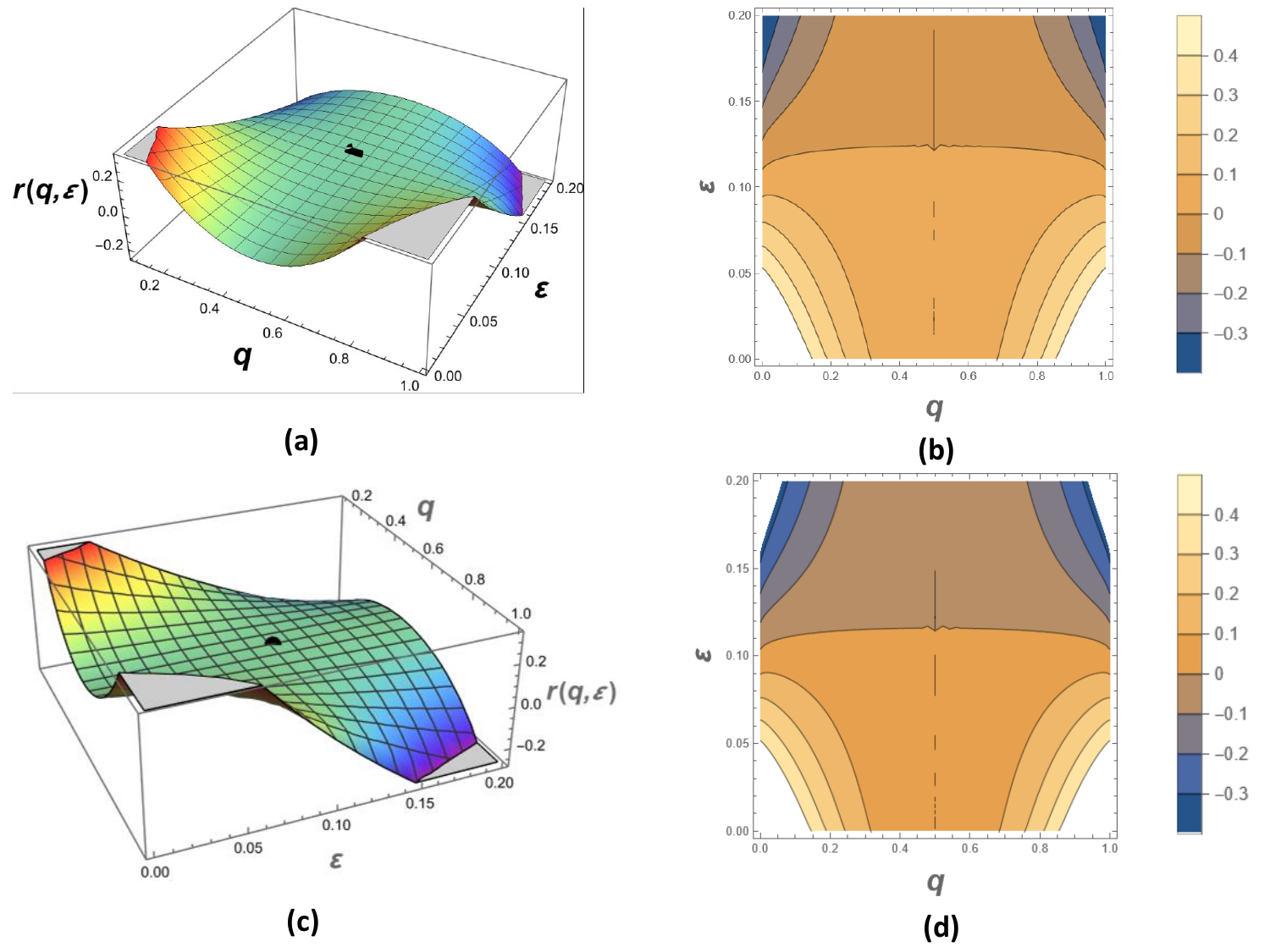} 
\par\end{centering}
\caption{\protect\label{fig:Lower=000020and=000020upper=000020bound=000020of=000020error=000020limit}(Color
online) Variation of secret key rate with bit-flip probability ($q$)
and QBER ($\mathcal{E}$): $(a)$ lower bound on the secret-key rate
of our protocol as a function of bit-flip probability and QBER, $(b)$
contour plot for lower bound error limit; QBER vs bit-flip probability,
$(c)$ upper bound on the secret-key rate of our protocol as a function
of bit-flip probability and QBER, and $(d)$ contour plot for upper
bound error limit; QBER vs bit-flip probability.}
\end{figure}

Now, we consider Bob's bit string, which he obtains from the measurement
result of his particle (system) $B$ in the state $|\Psi_{ABE}\rangle$.
Intuitively, there must be two equal possibilities for obtaining the
bit value $0$ and $1$ when considering Bob's bit string. Additionally,
if the conditional entropy of Bob's bit string is calculated provided
the noisy version of Alice's bit ($c$ value) string, then the error
and no-error probabilities will also be considered. So we would have,

\[
H(b)=1
\]
and 
\[
H(b|c)=h[q(1-\mathscr{\mathcal{E}})+(1-q)\mathcal{E}].
\]
Using these expressions, for an optimal choice of the parameter $q$,
we get the positive secret key if $\mathcal{E}\le0.124$ (refer to
Fig. \ref{fig:Lower=000020and=000020upper=000020bound=000020of=000020error=000020limit}
(a)). This tolerable limit for the error rate is under the classical
pre-processing, i.e., noise introduced by Alice.

Let us now determine
the upper bound of the secret key rate using Eq. \ref{eq:5}). As
before, the states of Eve, corresponding to the events where Alice
and Bob obtain the outcomes $(0,0),(0,0),(1,1)$, and $(1,1)$, are
given by $|\phi^{0,0}\rangle$, $|\phi^{0,+}\rangle$, $|\phi^{1,1}\rangle$,
and $|\phi^{1,-}\rangle$, respectively. Eve performs a von Neumann
measurement\footnote{This measurement is performed with respect to the
projectors along the states $\frac{1}{\sqrt{2}}\left(|\phi^{0,0}\rangle+|\phi^{1,1}\rangle\right)$,
$\frac{1}{\sqrt{2}}\left(|\phi^{0,0}\rangle-|\phi^{1,1}\rangle\right)$,
$\left(|\phi^{0,+}\rangle+|\phi^{1,-}\rangle\right)-\frac{1}{\sqrt{2}}\left(|\phi^{0,0}\rangle-|\phi^{1,1}\rangle\right)$,
and $\left(|\phi^{0,+}\rangle-|\phi^{1,-}\rangle\right)-\frac{1}{\sqrt{2}}\left(|\phi^{0,0}\rangle+|\phi^{1,1}\rangle\right)$.} to obtain her measurement outcome $e$. The conditional
entropy of Alice's noisy outcome, given Eve's measurement result,
is given by \cite{RGK_05,R08}

\[
\begin{array}{lcl}
H(c|e) & \ge & -\left(\mu_{1}+\mu_{2}\right)\left(h(\alpha)-k(\alpha,q)\right)\\
 & - & \left(\mu_{3}+\mu_{4}\right)\left(h(\beta)-k(\beta,q)\right)
\end{array}.
\]
Here, the parameters are defined as $\alpha=\frac{\mu_{1}}{\mu_{1}+\mu_{2}}$,
$\beta=\frac{\mu_{3}}{\mu_{3}+\mu_{4}}$, $k(m,n)=h\left(\frac{1}{2}\pm\frac{1}{2}\sqrt{1-16m\left(1-m\right)n\left(1-n\right)}\right)$,
$\mu_{1}+\mu_{2}=1-\mathcal{E}$, and $\mu_{3}+\mu_{4}=\mathcal{E}$.
By appropriately applying these conditions, Eq. (\ref{eq:5}) can
now be rewritten as follows:

\begin{equation}
\begin{array}{lcl}
r(a,b,e) & = & H(c|e)-H(c|b)\\
 & = & H(c|e)-[H(b|c)-H(b)]\\
 & \ge & -\left(1-\mathcal{E}\right)\left(h(\alpha)-k(\alpha,q)\right)-\mathcal{E}\left(h(\beta)-k(\beta,q)\right),\\
 & + & 1-h[q(1-\mathscr{\mathcal{E}})+(1-q)\mathcal{E}]
\end{array}\label{eq:new2}
\end{equation}

This allows us to compute the upper bound for the key rate by solving
$r(a,b,e)=0$. The solution yields an upper bound as  $\mathcal{E}\ge0.1125$, provided that optimal
value of $q$ is used (cf. Fig. \ref{fig:Lower=000020and=000020upper=000020bound=000020of=000020error=000020limit}
(c)). In the BB84 protocol, a one-way quantum channel is used for
a single transmission of the qubit sequence. However, the proposed
scheme requires three quantum channel transmissions due to its emphasis
on a greater quantum part and a reduced classical part. These three
transmissions involve distinct sequences prepared by Alice and Bob.
While the BB84 protocol employs one-way classical post-processing,
the same approach is retained in the proposed scheme. Given the three
quantum transmissions, it is essential to verify the security of each
sequence through information reconciliation. Since the three quantum
sequences are distinct, their security is analyzed separately. Our
analysis employs an information-theoretic approach, utilizing two-qubit
density operators to determine the upper and lower bounds of the secret
key rate based on the QBER, similar to the analysis used for the BB84
protocol. In the BB84 protocol, the tolerable QBER for the lower bound
(with classical pre-processing) and the upper bound of the key rate
are $\mathcal{E}\le0.124$ and $\mathcal{E}\ge0.146$, respectively.
In the proposed scheme, the corresponding QBER values are $\mathcal{E}\le0.124$
and $\mathcal{E}\ge0.1125$.

\section{Analysis of PNS attack\protect\label{sec:IV}}

\begin{figure}
\begin{centering}
\includegraphics[scale=0.3]{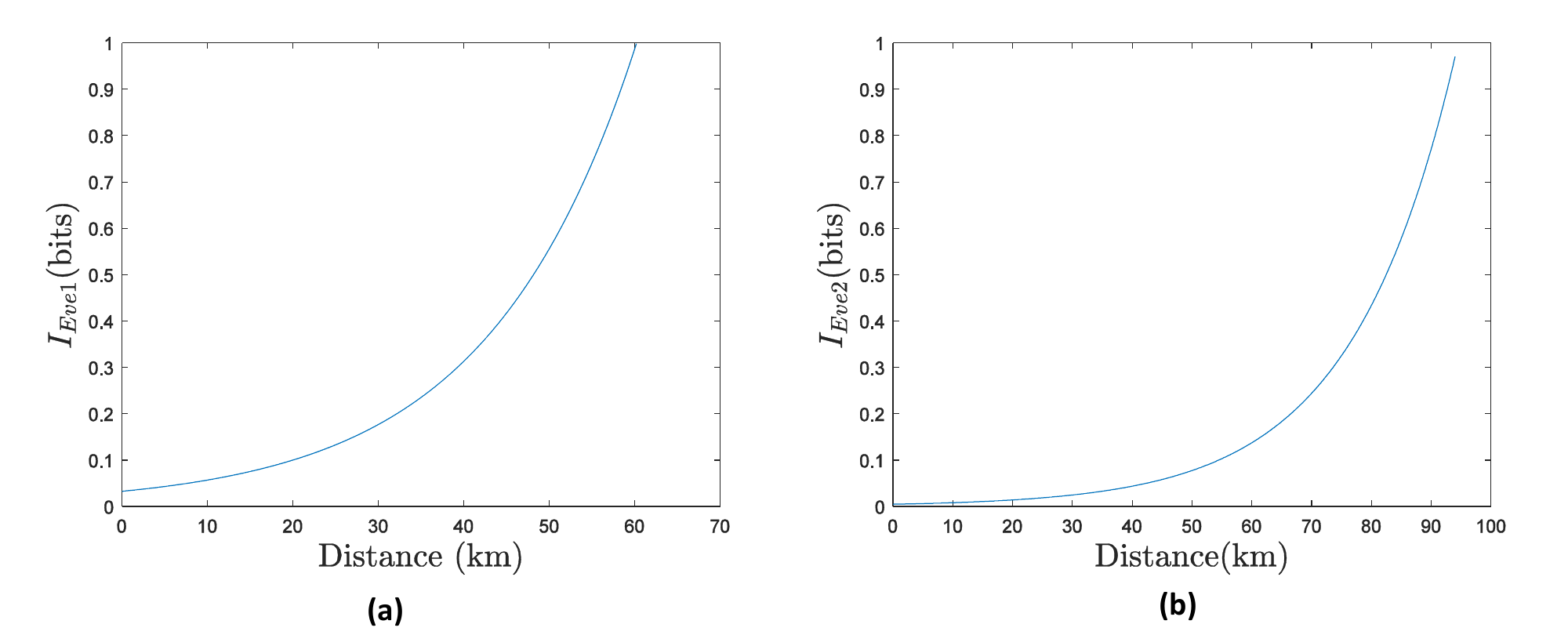} 
\par\end{centering}
\caption{\protect\label{fig:Eve's=000020information=000020vs=000020distance}(Color
online) Variation of Eve's information with distance to obtain critical
distance ($l_{c}$): $(a)$ Eve's information as a function of distance
to estimate the critical distance at which the attacker gains maximum
key information by the PNS attack on Protocol 1, $(b)$ Eve's information
as a function of distance to estimate the critical distance at which
the attacker gains maximum key information by the IRUD attack on Protocol
2.}
\end{figure}

We have already mentioned that our schemes can be realized using WCP
sources (see Eq. (\ref{eq:coherent})). However, a cryptographic scheme
based on WCP may face challenges due to the possibility of PNS and
similar attacks by an eavesdropper. Thus, we need to establish the
security of our schemes against different types of PNS attacks that
can be implemented by an eavesdropper (Eve) when $QBER=0$. The sub-protocols
in our scheme require Bob to announce basis information or a set of
non-orthogonal state information. Now, we note the following. 
\begin{enumerate}
\item In Protocol 1, Eve performs PNS attack with unlimited technological
power within the regime of laws of physics. Since the PNS attack in
its original form requires quantum memory, it is also called quantum
storage attack. Here, we consider the best scenario for Eve to attack.
The probability of getting an $n$ photon state is $p(n,\mu)=\frac{e^{-\mu}\mu^{n}}{n!}$,
where $\mu$ is the mean photon number. Alice and Bob know the channel
transmittance ($\eta$) and $\mu$. Both parties expect the probability
of detecting a non-zero photon in the absence of eavesdropping to
be $\underset{n\ge1}{\sum}p(n,\mu\eta)$, which is referred to as
the raw detection rate per pulse. In this context, sequence $S_{B2}$
is prepared independently with the same bit information as sequence
$S_{B1}$, but using a different basis. Eve targets only $S_{B1}$,
increasing her success probability of extracting information. Specifically,
we consider a scenario where Eve first performs a photon-number quantum
non-demolition (QND) measurement on $S_{B1}$ to count the number
of photons. She then blocks single-photon pulses and retains one photon
from multi-photon pulses. Subsequently, Eve sends the remaining photons
to Alice through a lossless channel\footnote{$\eta=10^{-\frac{\delta}{10}}$, and $\delta=\text{\ensuremath{\alpha l}{\rm [dB],}}$
where $\eta$ is the transmission in the fiber of length $l$, and
$\alpha$ is the loss in the fiber in dB/km. Since $\alpha$ and $l$
are non-negative quantities, $\delta$ is also non-negative. For the
minimum and maximum values of $\delta$ (i.e., for $\delta=0$ and
$\delta=\infty)$, we obtain $\eta=1$ and $\eta=0$, respectively.
Clearly $0\leq\eta\leq1$ quantifies the attenuation in a channel.
$\eta=1$ corresponds to a lossless channel with complete transmission
happens, and $\eta=0$ refers to an opaque channel with no transmission.} ($\eta=1$). However, Eve cannot arbitrarily block multi-photon pulses,
as the $QBER$ must remain zero. The extent to which Eve can block
multi-photon pulses depends on the channel transmittance $\eta$ between
Alice and Bob.

To execute this attack, the following condition must be satisfied
\cite{AGS04}:

\[
\underset{n\ge1}{\sum}p(n,\mu\eta)=t_{1}p(1,\mu)+\underset{n\ge2}{\sum}p(n,\mu),
\]

where $t_{1}$ is the fraction of single-photon pulses that reach
Bob. In the case of a powerful Eve, losses is such that $t_{1}=0$.
Under such conditions, Eve focuses exclusively on multi-photon pulses,
with the probability of this scenario being $\underset{n\ge2}{\sum}p(n,\mu)$.
The information gained by Eve is expressed as:

\[
I_{{\rm Eve1}}=\frac{0.5\times\underset{n\ge2}{\sum}p(n,\mu)}{0.75\times\underset{n\ge1}{\sum}p(n,\mu\eta)},
\]

The factor of 0.5 in the numerator originates from
the classical information disclosed by Bob during the generation of
the sifted key\footnote{For Protocol 1, a total of $0.625$\LyXThinSpace bits
of information is revealed to Eve, out of which only $0.5$\LyXThinSpace bits
are useful for generating the sifted key.}, while the factor of 0.75 in the denominator arises
from Alice\textquoteright s information related to non-empty pulses.
Specifically, 0.5 accounts for Bob\textquoteright s revealed classical
information, and the remaining 0.25 corresponds to Alice\textquoteright s
information gain in the absence of Bob\textquoteright s announcement.
Fig. \ref{fig:Eve's=000020information=000020vs=000020distance} (a)
depicts the variation of $I_{{\rm Eve}1}$ with distance $l$ for
an attenuation of $\alpha=0.25$ dB/km and a mean photon number $\mu=0.1$,
ensuring a fair comparison with the BB84 protocol ($\mu=0.1$). The
estimated critical attenuation is $\delta_{c}=15.05$ dB, corresponding
to a critical distance of $l_{c}=60.2$ km, beyond which the attacker
acquires full bit information under the PNS attack. Comparatively,
the critical distance for the BB84 protocol under similar conditions
is 52 km \cite{AGS04}, which is shorter than that of our protocol.
Additionally, the figure indicates that Eve gains almost no information
up to a distance of 30 km. This advantage stems from the utilization
of a higher amount of quantum signal or equivalently a higher amount
of quantum resources in the communication process. Furthermore, as
this protocol reveals less classical information than BB84, Eve's
ability to extract information after the classical announcement in
a collective attack scenario is reduced, thereby enhancing the security
of the final key.
\item In Protocol 2, Bob reveals the non-orthogonal state information. In
this scenario, Eve can execute a PNS-type attack, specifically referred
to as the intercept-resend with unambiguous discrimination (IRUD)
attack. In this attack, Eve begins by performing a photon-number QND
measurement to determine the number of photons in each pulse. She
discards all pulses containing less than three photons and proceeds
to measure the remaining pulses (those containing at least three photons)
using a measurement\footnote{The measurement $\mathcal{\mathscr{\mathcal{M}}}$ is any von Neumann
measurement that can discriminate the following four elements (states),
$|\Phi_{1}\rangle=\frac{1}{\sqrt{2}}\left(|000\rangle-|011\rangle\right),|\Phi_{2}\rangle=\frac{1}{2}\left(|101\rangle+|010\rangle+|100\rangle+|110\rangle\right),|\Phi_{3}\rangle=\frac{1}{\sqrt{2}}\left(|111\rangle-|001\rangle\right)$,
and $|\Phi_{4}\rangle=\frac{1}{2}\left(|101\rangle-|010\rangle-|100\rangle+|110\rangle\right)$\cite{SAR+04}.} operation $\mathcal{\mathscr{\mathcal{M}}}$. After obtaining a conclusive
result from $\mathcal{\mathscr{\mathcal{M}}}$, Eve prepares a new
photon state and forwards it to Bob. In this attack, it is assumed
that Eve operates using a lossless channel ($\eta=1$) and possesses
quantum memory. Eve performs this attack exclusively on sequence $S_{B1}$,
applying the same logic as the PNS attack used in Protocol 1. However,
Eve cannot arbitrarily discard pulses with fewer than three photons,
as the QBER must remain zero. For the attack to be successful, the
following condition must be satisfied:
\[
\underset{n\ge1}{\sum}p(n,\mu\eta)=t_{1}p(1,\mu)+t_{2}p(2,\mu)+\underset{n\ge3}{\sum}p(n,\mu),
\]
where $t_{1}$ and $t_{2}$ are the fractions of single-photon and
two-photon pulses, respectively, that reach Bob. The probabilities
for pulses containing more than three photons are negligible and can
be approximated as, $\underset{n\ge3}{\sum}p(n,\mu)\approx p(3,\mu)$.
For a powerful Eve, the losses are such that $t_{1}=0$ and $t_{2}=0$.
Under these conditions, Eve targets only three-photon pulses, which
occur with a probability of $p(3,\mu)$. The information gained by
Eve in this scenario is given as:
\[
I_{{\rm Eve2}}=\frac{I(3,\chi)p(3,\mu)}{\underset{n\ge1}{\sum}p(n,\mu\eta)},
\]
where $I(n,\chi)$ represents the maximum information Eve can extract
using $n$ photons in a single pulse. $\chi$ is the overlap of two
states within each set of non-orthogonal states announced by Bob\footnote{$I(n,\chi)=1-h(P,1-P)$ with $h(P,1-P)$ being the binary entropy
function and $P=\frac{1}{2}\left(1+\sqrt{1-\chi^{2n}}\right)$\cite{P97}.
In our case, the overlap, $\chi=\frac{1}{\sqrt{2}}.$}. The denominator represents the raw detection rate per pulse in the
absence of an eavesdropper, given a channel transmittance $\eta$
between Alice and Bob. For
a fair comparison with the SARG04 protocol ($\mu=0.2$), we also set
$\mu=0.2$. Assuming an attenuation factor of $\alpha=0.25$ dB/km
and $\mu=0.2$, we analyze the variation of Eve's information $(I_{Eve2})$
with distance to determine the critical attenuation (see Fig. \ref{fig:Eve's=000020information=000020vs=000020distance}
(b)). From Fig. \ref{fig:Eve's=000020information=000020vs=000020distance}
(b), the critical attenuation is found to be $\delta_{c}=23.75$ dB,
corresponding to a critical distance of $l_{c}=95$ km. In comparison,
under similar conditions, the critical distance for the SARG04 protocol
ranges from approximately 50 km to 100 km \cite{SAR+04}, which is
comparable to the critical distance achieved with our protocol. For
this attack, Eve gains almost no information up to 60 km. Additionally, our protocol reveals less classical
information, thereby offering better protection against Eve's collective
attack, which relies on the information disclosed in the classical
subprotocol.
\end{enumerate}

\section{Discussion\protect\label{sec:V}}

In this paper, we have proposed a new protocol for QKD and a variant
of it. The protocols consume more quantum resources compared to SARG04
or similar protocols but transmit less classical information over
the public channel, thereby reducing the probability of some side
channel attacks. Additionally, we conduct a rigorous security analysis
of the proposed protocols and calculate the tolerable error limit
for the upper and lower bounds of the secret key rate under a set
of collective attacks. It is shown that by applying a certain type
of classical pre-processing, the tolerable error limit can be increased.
The same is illustrated through the graphs. Before concluding, we
may emphasize on some important observations of our analysis. In the
seminal paper \cite{RGK_05}, the authors computed density operators
of Eve's final state in the six-state QKD protocol. Interestingly,
for our protocols, we obtained the same expressions for the density
operators describing Eve's final system, despite the fact that in
our Protocol 2 (1), neither Alice nor Bob (Alice never) discloses
the results of the measurements performed by them (her) in cases where
they (she) used different bases for preparation and measurement. The
reason for obtaining the same density operators for Eve's system is
that the terms that appear in the density matrix in cases of basis
mismatch happens, cancel each other. Additionally, we establish that
for the proposed protocols, the tolerable error limit of QBER $\mathcal{E}\le0.124$
for the lower bound of the key rate and $\mathcal{E}\ge0.1125$ for
the upper bound of the key rate if classical pre-processing is used.
In our case, the tolerable error limits are expected to decrease in
the absence of classical pre-processing.

In the practical implementation of cryptography, various types of
errors may occur during the transmission of qubits. Considering $QBER>0$,
Eve can attempt attack using partial cloning machines \cite{CI00,NG98,C00}.
Acin et al., have shown that legitimate users of SARG04 can tolerate
errors up to 15\% when Eve uses a best-known partial cloning machine.
They also found that this tolerable error limit is higher than that
for the BB84 protocol. In our case, for $QBER>0$, the tolerable error
limit is also computed to be 15\% (cf. Section \ref{sec:III}), which
is better than the BB84 protocol and its variants. We have also performed
a security-efficiency trade-off analysis for the proposed schemes
and compared their efficiency with the SARG04 protocol, as detailed
in Appendix F.

\subsection*{Acknowledgment: }

Authors acknowledge support from the QUEST scheme of the Interdisciplinary
Cyber-Physical Systems (ICPS) program of the Department of Science
and Technology (DST), India, Grant No.: DST/ICPS/QuST/Theme-1/2019/14
(Q80). They also thank Kishore Thapliyal and Sandeep Mishra for their
interest and useful technical feedback on this work.

\section*{Availability of data and materials}

No additional data is needed for this work.

\section*{Competing interests}

The authors declare that they have no competing interests.

\section*{Authors' contribution}

AD and AP conceptualized the problem. AD performed most of the calculations
and prepared the first draft of the manuscript. AP supervised the
work, checked the calculations, and prepared the final draft of the
paper.

\bibliographystyle{unsrt}
\bibliography{new-QKD}

\appendix

\section*{Appendix A \protect\label{sec:Appendix-A}}

It is already discussed in the main text that the Bell states are,
$|\Phi^{\pm}\rangle\text{=\ensuremath{\frac{1}{\sqrt{2}}(|00\rangle\pm|11\rangle})}$
and $|\Psi^{\pm}\rangle$$=\ensuremath{\frac{1}{\sqrt{2}}(|01\rangle\pm|10\rangle})$.
We can also express the Bell states in the diagonal basis as $|\Phi^{+}\rangle=\frac{1}{\sqrt{2}}\left(|++\rangle+|--\rangle\right)$,
$|\Phi^{-}\rangle=\frac{1}{\sqrt{2}}\left(|+-\rangle+|-+\rangle\right)$,
$|\Psi^{+}\rangle=\frac{1}{\sqrt{2}}\left(|++\rangle-|--\rangle\right)$,
and $|\Psi^{-}\rangle=\frac{1}{\sqrt{2}}\left(|-+\rangle-|+-\rangle\right)$.
We may now write from Eq. (\ref{eq:1}),

\begin{equation}
\begin{array}{lcl}
\mu_{2} & = & \langle\Phi^{-}|\rho|\Phi^{-}\rangle\\
 & = & \frac{1}{2}\left(\langle+-|\rho|+-\rangle+\langle+-|\rho|-+\rangle+\langle-+|\rho|+-\rangle+\langle-+|\rho|-+\rangle\right),
\end{array}\label{eq:Mu2}
\end{equation}

\begin{equation}
\begin{array}{lcl}
\mu_{4} & = & \langle\Psi^{-}|\rho|\Psi^{-}\rangle\\
 & = & \frac{1}{2}\left(\langle+-|\rho|+-\rangle-\langle+-|\rho|-+\rangle-\langle-+|\rho|+-\rangle+\langle-+|\rho|-+\rangle\right),
\end{array}\label{eq:Mu4}
\end{equation}

\begin{equation}
\begin{array}{lcl}
\mu_{3} & = & \langle\Psi^{+}|\rho|\Psi^{+}\rangle\\
 & = & \frac{1}{2}\left(\langle01|\rho|01\rangle+\langle01|\rho|10\rangle+\langle10|\rho|01\rangle+\langle10|\rho|10\rangle\right),
\end{array}\label{eq:Mu3}
\end{equation}
and

\begin{equation}
\begin{array}{lcl}
\mu_{4} & = & \mathopen{\langle}{\Psi^{-}|\rho|\Psi^{-}}\mathclose{\rangle}\\
 & = & \frac{1}{2}\left(\langle01|\rho|01\rangle-\langle01|\rho|10\rangle-\langle10|\rho|01\rangle+\langle10|\rho|10\rangle\right).
\end{array}\label{eq:Mu4-1}
\end{equation}
We consider $\mathcal{E}$ to be a symmetric error, and therefore,
the following relationships are to be valid,

\begin{equation}
\begin{array}{lcl}
\langle00|\rho|00\rangle+\langle11|\rho|11\rangle & = & 1-\mathcal{E},\\
\langle++|\rho|++\rangle+\langle--|\rho|--\rangle & = & 1-\mathcal{E},\\
\langle01|\rho|01\rangle+\langle10|\rho|10\rangle & = & \mathcal{E},\\
\langle+-|\rho|+-\rangle+\langle-+|\rho|-+\rangle & = & \mathcal{E}.
\end{array}\label{eq:ErrorRelation}
\end{equation}
Inserting these Eqs. (\ref{eq:Mu2}), (\ref{eq:Mu4}), (\ref{eq:Mu3}),
and (\ref{eq:Mu4-1}) in the relation (\ref{eq:ErrorRelation}), we
obtain

\[
\begin{array}{lcl}
\mu_{2}+\mu_{4} & = & \left(\langle+-|\rho|+-\rangle+\langle-+|\rho|-+\rangle\right)=\mathcal{E}\\
\mu_{2} & = & \mathcal{E}-\mu_{4}
\end{array},
\]
and

\textbf{ 
\[
\begin{array}{lcl}
\mu_{3}+\mu_{4} & = & \left(\langle01|\rho|01\rangle+\langle10|\rho|10\rangle\right)=\mathcal{E}\\
\mu_{3} & = & \mathcal{E}-\mu_{4}.
\end{array}.
\]
}Now, total probability must satisfy $\mu_{1}+\mu_{2}+\mu_{3}+\mu_{4}=1$.
By employing the aforementioned connection with the preceding outcomes,
we obtain, $\mu_{1}=1-2\mathcal{E}+\mu_{4}$ and $\mu_{2}=\mu_{3}=\mathcal{E}-\mu_{4}$.

\section*{Appendix B \protect\label{sec:Appendix-B}}

We use these relations to compute the key rate. The conditional probability,
$Pr\left(B=|i\rangle\left|A=|j\rangle\right.\right)=\frac{Pr(B=|i\rangle,A=|j\rangle)}{Pr(A=|j\rangle)}$
and conditional entropy,

\begin{equation}
H(B|A)=-\sum_{j}Pr(A=|j\rangle)\sum_{i}Pr(B=|i\rangle|A=|j\rangle)\log_{2}Pr(B=|i\rangle|A=|j\rangle),\label{eq:11conditional=000020entropy}
\end{equation}

\[
\begin{array}{lcl}
Pr\left(B=|0\rangle|A=|0\rangle\right)=Pr\left(B=|1\rangle|A=|1\rangle\right) & = & \frac{\mu_{1}+\mu_{2}}{2},\\
Pr\left(B=|1\rangle|A=|0\rangle\right)=Pr\left(B=|0\rangle|A=|1\rangle\right) & = & \frac{\mu_{3}+\mu_{4}}{2},\\
Pr\left(B=|+\rangle|A=|+\rangle\right)=Pr\left(B=|-\rangle|A=|-\rangle\right) & = & \frac{\mu_{1}+\mu_{3}}{2},\\
Pr\left(B=|-\rangle|A=|+\rangle\right)=Pr\left(B=|+\rangle|A=|-\rangle\right) & = & \frac{\mu_{2}+\mu_{4}}{2},
\end{array}
\]

\[
Pr\left(B=|m\rangle|A=|n\rangle\right)=Pr\left(B=|n\rangle|A=|m\rangle\right)=\frac{1}{4},
\]

\[
Pr\left(B=|m\rangle\right)=Pr\left(B=|n\rangle\right)=\frac{1}{4},
\]

\[
Pr\left(A=|m\rangle\right)=Pr\left(A=|n\rangle\right)=\frac{1}{4},
\]
here $m\neq n$ and $m\in\{|0\rangle,|1\rangle\}$ and $n\in\{|+\rangle,|-\rangle,\}$.
Now, using Eq. (\ref{eq:11conditional=000020entropy}) we can obtain

\[
\begin{array}{lcl}
H(B|A) & = & -\frac{\mu_{1}+\mu_{2}}{4}\log_{2}\frac{\mu_{1}+\mu_{2}}{2}-\frac{\mu_{3}+\mu_{4}}{4}\log_{2}\frac{\mu_{3}+\mu_{4}}{2}\\
 & - & \frac{\mu_{1}+\mu_{3}}{4}\log_{2}\frac{\mu_{1}+\mu_{3}}{2}-\frac{\mu_{2}+\mu_{4}}{4}\log_{2}\frac{\mu_{2}+\mu_{4}}{2}-\frac{1}{2}\log_{2}\frac{1}{4}\\
 & = & -\frac{1-\mathcal{E}}{2}\log_{2}\frac{1-\mathcal{E}}{2}-\frac{\mathcal{E}}{2}\log_{2}\frac{\mathcal{E}}{2}+1,
\end{array}
\]
and 
\[
\begin{array}{lcl}
H(B) & = & -4\times\frac{1}{4}\log_{2}\frac{1}{4}\\
 & = & 2
\end{array}.
\]

Therefore,

\[
\begin{array}{lcl}
I(A:B) & = & H(B)-H(B|A)\\
 & = & 1+\frac{1-\mathcal{E}}{2}\log_{2}\frac{1-\mathcal{E}}{2}+\frac{\mathcal{E}}{2}\log_{2}\frac{\mathcal{E}}{2},
\end{array}
\]
using the secret key rate we have,

\[
\begin{array}{lcl}
r & = & I(A:B)-H(V)\\
 & = & 1+\frac{1-\mathcal{E}}{2}\log_{2}\frac{1-\mathcal{E}}{2}+\frac{\mathcal{E}}{2}\log_{2}\frac{\mathcal{E}}{2}-2h(\mathcal{E}).
\end{array}
\]

\section*{Appendix C\protect\label{sec:Appendix-C}}

Here, we calculate the key rate equation after the key-sifting subprotocol
by the both parties which means that the probability in acceptable
condition will be considered.

\[
\begin{array}{lcl}
Pr\left(b=0|a=0\right)=Pr\left(b=1|a=1\right) & = & \frac{1}{6}+\frac{2\mu_{1}+\mu_{2}+\mu_{3}}{3},\\
Pr\left(b=0|a=1\right)=Pr\left(b=1|a=0\right) & = & \frac{1}{6}+\frac{\mu_{2}+\mu_{3}+2\mu_{4}}{3},
\end{array}
\]

\[
\begin{array}{lcl}
Pr\left(b=0\right)=Pr\left(b=1\right) & = & \frac{1}{2},\end{array}
\]
in the similar approach for Eq. (\ref{eq:11conditional=000020entropy})
we have, 
\[
\begin{array}{lcl}
H(b|a) & = & -\frac{1+4\mu_{1}+2\mu_{2}+2\mu_{3}}{6}\log_{2}\frac{1+4\mu_{1}+2\mu_{2}+2\mu_{3}}{6}-\frac{1+2\mu_{2}+2\mu_{3}+4\mu_{4}}{6}\log_{2}\frac{1+2\mu_{2}+2\mu_{3}+4\mu_{4}}{6}\\
 & = & -\frac{5-4\mathcal{E}}{6}\log_{2}\frac{5-4\mathcal{E}}{6}-\frac{1+4\mathcal{E}}{6}\log_{2}\frac{1+4\mathcal{E}}{6}\\
 & = & h\left(\frac{1}{6}+\frac{2\mathcal{E}}{3}\right),
\end{array}
\]

\[
\begin{array}{lcl}
I(a|b) & = & H(b)-H(b|a)\\
 & = & -\frac{1}{2}\log_{2}\frac{1}{2}-h(b|a)\\
 & = & 1-h\left(\frac{1}{6}+\frac{2\mathcal{E}}{3}\right),
\end{array}
\]
so the final expression for secret key rate,

\[
\begin{array}{lcl}
r & = & I(a|b)-H(V)\\
 & = & 1-h\left(\frac{1}{6}+\frac{2\mathcal{E}}{3}\right)-2h(\mathcal{E}).
\end{array}
\]

\section*{Appendix D\protect\label{sec:Appendix-D}}

Alice and Bob are evaluating the security threshold of the particle
sequences, denoted as $S_{A}$ and $S_{B1}$, when they use the same
basis for preparing or measuring the states. The absence or presence
of errors can be defined when they measure the states $|\Phi^{\pm}\rangle$
or $|\Psi^{\pm}\rangle)$ using the computational basis. Similarly,
there will be an absence or presence of errors when both Alice and
Bob measure the states $|\Phi^{+}\rangle$ and $|\Psi^{+}\rangle$
or $|\Phi^{-}\rangle$ and $|\Psi^{-}\rangle$ with the diagonal basis.
These scenarios lead to four cases that we need to consider. Furthermore,
we can conclude that the state $\rho$ can be measured with equal
probability by both Alice and Bob using both the computational and
diagonal bases. To enhance comprehension, let us begin with an illustrative
example. Consider the scenario where two parties measure the Bell
states $|\Phi^{+}\rangle$ and $|\Psi^{+}\rangle$ in the computational
basis. The resulting probabilities of encountering no error and error
are $\frac{\mu_{1}}{2}$ and $\frac{\mu_{3}}{2}$, respectively\footnote{To keep things simple, we assume that both participants measure all
particles using the same basis (either both use the computational
basis or both use diagonal basis) during the error checking step.}. Consequently, the probabilities of experiencing no error and error
when measuring the states $|\Phi^{+}\rangle$ and $|\Psi^{+}\rangle$
are $\frac{1-\mathcal{E}}{2}$ and $\frac{\mathcal{E}}{2}$, respectively\footnote{In this context, the factor of 2 emerges because we evenly distribute
the total error-checking qubits between computational and diagonal
basis measurements.}. It is evident that the probability of encountering no error and
error when measuring the states $|\Phi^{+}\rangle$ and $|\Psi^{+}\rangle$
in the computational basis, considering the total number of qubits,
can be expressed as $\frac{\frac{\mu_{1}}{2}}{\frac{1-\mathcal{E}}{2}}$
and $\frac{\frac{\mu_{3}}{2}}{\frac{\mathcal{E}}{2}}$, which simplifies
to$\frac{\mu_{1}}{1-\mathcal{E}}$ and $\frac{\mu_{3}}{\mathcal{E}}$,
respectively. Similarly, when measuring the states $|\Phi^{-}\rangle$
and $|\Psi^{-}\rangle$ in the computational basis, the probabilities
of encountering no error and error can be expressed as $\frac{\mu_{2}}{1-\mathcal{E}}$
and $\frac{\mu_{4}}{\mathcal{E}}$. Moving on to the diagonal basis,
the probabilities of encountering no error and error when measuring
the states $|\Phi^{+}\rangle$ and $|\Psi^{+}\rangle$ are $\frac{\mu_{1}}{1-\mathcal{E}}$
and $\frac{\mu_{3}}{1-\mathcal{E}}$, respectively. Similarly, for
the states $|\Phi^{-}\rangle$ and |$\Psi^{-}\rangle$, the probabilities
are $\frac{\mu_{2}}{\mathcal{E}}$ and $\frac{\mu_{4}}{\mathcal{E}}$.
In a scenario where no errors occur, one can calculate the entropy
as follows:

\[
\begin{array}{lcl}
{\rm H}_{{\rm no-error}} & = & -\frac{1}{2}\left[\frac{\mu_{1}}{1-\mathcal{E}}\log_{2}\frac{\mu_{1}}{1-\mathcal{E}}+\frac{\mu_{2}}{1-\mathcal{E}}\log_{2}\frac{\mu_{2}}{1-\mathcal{E}}+\frac{\mu_{1}}{1-\mathcal{E}}\log_{2}\frac{\mu_{1}}{1-\mathcal{E}}+\frac{\mu_{3}}{1-\mathcal{E}}\log_{2}\frac{\mu_{3}}{1-\mathcal{E}}\right]\\
 & = & -\frac{1}{2}\left[\frac{2\mu_{1}}{1-\mathcal{E}}\log_{2}\frac{\mu_{1}}{1-\mathcal{E}}+\frac{2\mu_{2}}{1-\mathcal{E}}\log_{2}\frac{\mu_{2}}{1-\mathcal{E}}\right]\,\,\,\,{\rm as}\,\,\mu_{2}=\mu_{3}\\
 & = & -\left[\frac{\left(1-2\mathcal{E}+\mu_{4}\right)}{1-\mathcal{E}}\log_{2}\frac{\left(1-2\mathcal{E}+\mu_{4}\right)}{1-\mathcal{E}}+\frac{\left(\mathcal{E}-\mu_{4}\right)}{1-\mathcal{E}}\log_{2}\frac{\left(\mathcal{E}-\mu_{4}\right)}{1-\mathcal{E}}\right]\,\,\,\,({\rm using\,the\,results\,of\,Appendix\,A})\\
 & = & h\left(\frac{1-2\mathcal{E}+\mu_{4}}{1-\mathcal{E}}\right),
\end{array}
\]

and in presence of error the entropy can be computed as,

\[
\begin{array}{lcl}
{\rm H_{error}} & = & -\frac{1}{2}\left[\frac{\mu_{3}}{\mathcal{E}}\log_{2}\frac{\mu_{3}}{\mathcal{E}}+\frac{\mu_{4}}{\mathcal{E}}\log_{2}\frac{\mu_{4}}{\mathcal{E}}+\frac{\mu_{2}}{\mathcal{E}}\log_{2}\frac{\mu_{2}}{\mathcal{E}}+\frac{\mu_{4}}{\mathcal{E}}\log_{2}\frac{\mu_{4}}{\mathcal{E}}\right]\\
 & = & -\frac{1}{2}\left[\frac{2\mu_{2}}{\mathcal{E}}\log_{2}\frac{2\mu_{2}}{\mathcal{E}}+\frac{2\mu_{4}}{\mathcal{E}}\log_{2}\frac{2\mu_{4}}{\mathcal{E}}\right]\,\,{\rm as}\,\mu_{2}=\mu_{3}\\
 & = & -\left[\frac{\left(\mathcal{E}-\mu_{4}\right)}{\mathcal{E}}\log_{2}\frac{\left(\mathcal{E}-\mu_{4}\right)}{\mathcal{E}}+\frac{\mu_{4}}{\mathcal{E}}\log_{2}\frac{\mu_{4}}{\mathcal{E}}\right]\,\,({\rm using\,the\,results\,of\,Appendix\,A})\\
 & = & h\left(\frac{\mathcal{E}-\mu_{4}}{\mathcal{E}}\right).
\end{array}
\]

After conducting statistical averaging for both no error and error
scenarios, we acquire,

\[
\begin{array}{l}
\left(1-\mathcal{E}\right){\rm H}_{{\rm no-error}}+\mathcal{E}H_{error}\\
=\left(1-\mathcal{E}\right)h\left(\frac{1-2\mathcal{E}+\mu_{4}}{1-\mathcal{E}}\right)+\mathcal{E}h\left(\frac{\mathcal{E}-\mu_{4}}{\mathcal{E}}\right)\\
=-\left(1-\mathcal{E}\right)\left[\frac{\left(1-2\mathcal{E}+\mu_{4}\right)}{1-\mathcal{E}}\log_{2}\frac{\left(1-2\mathcal{E}+\mu_{4}\right)}{1-\mathcal{E}}+\frac{\left(\mathcal{E}-\mu_{4}\right)}{1-\mathcal{E}}\log_{2}\frac{\left(\mathcal{E}-\mu_{4}\right)}{1-\mathcal{E}}\right]\\
-\mathcal{E}\left[\frac{\left(\mathcal{E}-\mu_{4}\right)}{\mathcal{E}}\log_{2}\frac{\left(\mathcal{E}-\mu_{4}\right)}{\mathcal{E}}+\frac{\mu_{4}}{\mathcal{E}}\log_{2}\frac{\mu_{4}}{\mathcal{E}}\right]\\
=-\left[\left(1-2\mathcal{E}+\mu_{4}\right)\log_{2}\left(1-2\mathcal{E}+\mu_{4}\right)+\left(\mathcal{E}-\mu_{4}\right)\log_{2}\left(\mathcal{E}-\mu_{4}\right)-\left(1-2\mathcal{E}+\mu_{4}\right)\log_{2}\left(1-\mathcal{E}\right)-\left(\mathcal{E}-\mu_{4}\right)\log_{2}\left(1-\mathcal{E}\right)\right]\\
-\left[\left(\mathcal{E}-\mu_{4}\right)\log_{2}\left(\mathcal{E}-\mu_{4}\right)+\mu_{4}\log_{2}\mu_{4}-\left(\mathcal{E}-\mu_{4}\right)\log_{2}\mathcal{E}-\mu_{4}\log_{2}\mathcal{E}\right]\\
=-\left[\left(1-2\mathcal{E}+\mu_{4}\right)\log_{2}\left(1-2\mathcal{E}+\mu_{4}\right)+2\left(\mathcal{E}-\mu_{4}\right)\log_{2}\left(\mathcal{E}-\mu_{4}\right)+\mu_{4}\log_{2}\mu_{4}\right]\\
+\left(1-2\mathcal{E}+\mu_{4}+\mathcal{E}-\mu_{4}\right)\log_{2}\left(1-\mathcal{E}\right)+\left(\mathcal{E}-\mu_{4}+\mu_{4}\right)\log_{2}\mathcal{E}\\
=H(V)+\left(1-\mathcal{E}\right)\log_{2}\left(1-\mathcal{E}\right)+\mathcal{E}\log_{2}\mathcal{E}\\
=H(V)-h(\mathcal{E}).
\end{array}
\]

\section*{Appendix E\protect\label{sec:Appendix-E}}

In this appendix, we provide a detailed explanation of the mathematical
processes involved in obtaining Eq. (\ref{eq:8}) from Eq. (\ref{eq:Composite_state_with_Eve})
in Section \ref{sec:III}. To begin with, let us focus on a scenario
in which both Alice and Bob perform measurements on their qubits using
the $Z$ basis (similar outcomes are observed for the $X$ basis as
well).

\[
\begin{array}{lcl}
|\Psi\rangle_{ABE} & := & \stackrel[i=1]{4}{\sum}\sqrt{\mu_{i}}|\varphi_{i}\rangle_{AB}\otimes|\varepsilon_{i}\rangle_{E}\\
 & = & \frac{1}{2\sqrt{2}}\left[\sqrt{\mu_{1}}\left(|00\rangle+|11\rangle\right)\otimes|\varepsilon_{1}\rangle+\sqrt{\mu_{2}}\left(|00\rangle-|11\rangle\right)\otimes|\varepsilon_{2}\rangle\right.\\
 & + & \left.\sqrt{\mu_{3}}\left(|01\rangle+|10\rangle\right)\otimes|\varepsilon_{3}\rangle+\sqrt{\mu_{4}}\left(|01\rangle-|10\rangle\right)\otimes|\varepsilon_{4}\rangle\right]_{ABE}\\
 & = & \frac{1}{2\sqrt{2}}\left[|00\rangle\left(\sqrt{\mu_{1}}|\varepsilon_{1}\rangle+\sqrt{\mu_{2}}|\varepsilon_{2}\rangle\right)+|11\rangle\left(\sqrt{\mu_{1}}|\varepsilon_{1}\rangle-\sqrt{\mu_{2}}|\varepsilon_{2}\rangle\right)\right.\\
 & + & \left.|01\rangle\left(\sqrt{\mu_{3}}|\varepsilon_{3}\rangle+\sqrt{\mu_{4}}|\varepsilon_{4}\rangle\right)+|10\rangle\left(\sqrt{\mu_{3}}|\varepsilon_{3}\rangle-\sqrt{\mu_{4}}|\varepsilon_{4}\rangle\right)\right]_{ABE}\\
 & = & \frac{1}{2}\left[|00\rangle\otimes|\phi^{0,0}\rangle+|11\rangle\otimes|\phi^{1,1}\rangle+|01\rangle\otimes|\phi^{0,1}\rangle+|10\rangle\otimes{|\phi^{1,0}}\mathclose{\rangle}\right]_{ABE}
\end{array}.
\]
Next, consider the scenario in which Alice and Bob measure their qubits
using the $Z$ and $X$ bases, respectively\footnote{It is important to note that the same outcome occurs when Alice and
Bob opt for the $X$ and $Z$ bases, respectively.},

\[
\begin{array}{lcl}
|\Psi\rangle_{ABE} & := & \stackrel[i=1]{4}{\sum}\sqrt{\mu_{i}}|\varphi_{i}\rangle_{AB}\otimes|\varepsilon_{i}\rangle_{E}\\
 & = & \frac{1}{2\sqrt{2}}\left[\sqrt{\mu_{1}}\left(|00\rangle+|11\rangle\right)\otimes|\varepsilon_{1}\rangle+\sqrt{\mu_{2}}\left(|00\rangle-|11\rangle\right)\otimes|\varepsilon_{2}\rangle\right.\\
 & + & \left.\sqrt{\mu_{3}}\left(|01\rangle+|10\rangle\right)\otimes|\varepsilon_{3}\rangle+\sqrt{\mu_{4}}\left(|01\rangle-|10\rangle\right)\otimes|\varepsilon_{4}\rangle\right]_{ABE}\\
 & = & \frac{1}{4}\left[\sqrt{\mu_{1}}\left\{ |0\rangle\left(|+\rangle+|-\rangle\right)+|1\rangle\left(|+\rangle-|-\rangle\right)\right\} |\varepsilon_{1}\rangle\right.\\
 & + & \sqrt{\mu_{2}}\left\{ |0\rangle\left(|+\rangle+|-\rangle\right)-|1\rangle\left(|+\rangle-|-\rangle\right)\right\} |\varepsilon_{2}\rangle\\
 & + & \sqrt{\mu_{3}}\left\{ |0\rangle\left(|+\rangle-|-\rangle\right)+|1\rangle\left(|+\rangle+|-\rangle\right)\right\} |\varepsilon_{3}\rangle\\
 & + & \left.\sqrt{\mu_{4}}\left\{ |0\rangle\left(|+\rangle-|-\rangle\right)-|1\rangle\left(|+\rangle+|-\rangle\right)\right\} |\varepsilon_{4}\rangle\right]_{ABE}\\
 & = & \frac{1}{4}\left[|0+\rangle\left\{ \sqrt{\mu_{1}}|\varepsilon_{1}\rangle+\sqrt{\mu_{2}}|\varepsilon_{2}\rangle+\sqrt{\mu_{3}}|\varepsilon_{3}\rangle+\sqrt{\mu_{4}}|\varepsilon_{4}\rangle\right\} \right.\\
 & + & |0-\rangle\left\{ \sqrt{\mu_{1}}|\varepsilon_{1}\rangle+\sqrt{\mu_{2}}|\varepsilon_{2}\rangle-\sqrt{\mu_{3}}|\varepsilon_{3}\rangle-\sqrt{\mu_{4}}|\varepsilon_{4}\rangle\right\} \\
 & + & |1+\rangle\left\{ \sqrt{\mu_{1}}|\varepsilon_{1}\rangle-\sqrt{\mu_{2}}|\varepsilon_{2}\rangle+\sqrt{\mu_{3}}|\varepsilon_{3}\rangle-\sqrt{\mu_{4}}|\varepsilon_{4}\rangle\right\} \\
 & + & \left.|1-\rangle\left\{ -\sqrt{\mu_{1}}|\varepsilon_{1}\rangle+\sqrt{\mu_{2}}|\varepsilon_{2}\rangle+\sqrt{\mu_{3}}|\varepsilon_{3}\rangle-\sqrt{\mu_{4}}|\varepsilon_{4}\rangle\right\} \right]_{ABE}\\
 & = & \frac{1}{2}\left[|0+\rangle|\phi^{0,+}\rangle+|0-\rangle|\phi^{0,-}\rangle+|1+\rangle|\phi^{1,+}\rangle+|1-\rangle|\phi^{1,-}\rangle\right]_{ABE}
\end{array},
\]
In the main text, we have provided the details of Eve's initial state,
denoted as $\varepsilon_{E}$, as well as Eve's state $|\phi^{\mathcal{A},\mathcal{B}}\rangle$
after Alice and Bob's measurement. Our focus is solely on the instances
that Alice and Bob accept after the classical pre-processing stage.
Following normalization, we examine the density operator of Eve's
system specifically when Alice obtains the outcome $0$,

\[
\begin{array}{lcl}
\sigma_{E}^{0} & = & \frac{1}{2}\left[|\phi^{0,0}\rangle\langle\phi^{0,0}|+|\phi^{0,1}\rangle\langle\phi^{0,1}|\right]+\frac{1}{2}\left[|\phi^{0,+}\rangle\langle\phi^{0,+}|+|\phi^{0,-}\rangle\langle\phi^{0,-}|\right]\\
 & = & \frac{1}{2}\left(P_{|\phi^{0,0}\rangle}+P_{|\phi^{0,1}\rangle}\right)+\frac{1}{2}\left(P_{|\phi^{0,+}\rangle}+P_{|\phi^{0,-}\rangle}\right)
\end{array},
\]
and when Alice obtains the outcome $1$ is,

\[
\begin{array}{lcl}
\sigma_{E}^{1} & = & \frac{1}{2}\left[|\phi^{1,0}\rangle\langle\phi^{1,0}|+|\phi^{1,1}\rangle\langle\phi^{1,1}|\right]+\frac{1}{2}\left[|\phi^{1,+}\rangle\langle\phi^{1,+}|+|\phi^{1,-}\rangle\langle\phi^{1,-}|\right]\\
 & = & \frac{1}{2}\left(P_{|\phi^{1,0}\rangle}+P_{|\phi^{1,1}\rangle}\right)+\frac{1}{2}\left(P_{|\phi^{1,+}\rangle}+P_{|\phi^{1,-}\rangle}\right)
\end{array}.
\]

\section*{Appendix F \protect\label{sec:Appendix-F}}

\emph{Security-efficiency trade-off for our protocols: }In 2000, Cabello
\cite{C2000} introduced a measure of the efficiency of quantum communication
protocols as $\eta=\frac{b_{s}}{q_{t}+b_{t}}$, where $b_{s}$ is
the number of secret bits exchanged by the protocol, $q_{t}$ is the
number of qubits interchanged via the quantum channel in each step
of the protocol, and $b_{t}$ is the classical bit information exchanged
between Alice and Bob via the classical channel\footnote{The classical bit which is used for detecting eavesdropping is neglected
here.}. Considering Protocol 2, an inherent error arises
during its execution. If Alice's initial state is $|0\rangle$, and
Bob's measurement outcomes are $|0\rangle$ or $|-\rangle$, no error
occurs. However, if Bob's outcome is $|+\rangle$, an inherent error
is introduced. The probabilities of Bob obtaining $|0\rangle$, $|-\rangle$,
and $|+\rangle$ are $\frac{1}{8}$, $\frac{1}{16}$, and $\frac{1}{16}$,
respectively (see Table \ref{tab:Main=000020table=000020for=000020protocol=0000201=000020and=0000202}).
In this case, when the measurement result is determined as $|0\rangle$,
the error probability corresponding to Bob\textquoteright s measurement
outcomes $|0\rangle$ and $|+\rangle$ is calculated as $\left(\frac{1}{64}\right)/\left(\frac{1}{64}+\frac{1}{8}\right)=\frac{1}{9}$.
A similar scenario applies when Alice\textquoteright s initial state
is $|1\rangle$, $|+\rangle$ or $|-\rangle$. For instance, Alice\textquoteright s
initial state $|1\rangle$, no error occurs when Bob\textquoteright s
measurement results are $\left\{ 1,+\right\} $, whereas an error
arises when the outcome is $|-\rangle$. The same probability calculations
hold for the remaining cases of Alice\textquoteright s initial states.
Finally, the secret key bits ($b_{s}$) are transformed as follows:

\[
\begin{array}{lcl}
b_{s} & = & \left[\left(\frac{1}{16}+\frac{1}{32}+\frac{1}{64}\right)+\left(\frac{1}{8}+\frac{1}{64}\right)\left(1-h(\frac{1}{9})\right)\right]\times4\\
 & = & 0.72
\end{array}.
\]
For the sifting subprotocol of the second QKD protocol (Protocol 2),
the values of the essential parameters are $b_{s}=0.72$, $q_{t}=3$
and $b_{t}=0.75$, resulting in an efficiency of $\eta=0.192$. For
the sifting subprotocol of our Protocol 1, the values of the essential
parameter are $b_{s}=0.75$, $q_{t}=3$ and $b_{t}=0.625$, giving
an efficiency of $\eta=0.2069$ with no inherent error. In this specific
sifting condition, the basis information will be revealed at the end
of the protocol, which may increase the chance of a PNS attack by
a powerful Eve. To apply the more efficient Protocol 2, one must consider
the inherent error probability value of $0.0625$, and the exchange
of classical information is also more than Protocol 1. We want to
stress that the Protocol 2 is more robust against PNS attack because
Alice and Bob do not reveal the basis information; instead, two non-orthogonal
state information values are announced for the sifting process ($M$
value). Our two protocols are more efficient than the SARG04 protocol\footnote{The values of essential parameters for SARG04 protocol are $b_{s}=0.25$,
$q_{t}=1$ and $b_{t}=1$, resulting in an efficiency of $\eta=0.125$.}($\eta=0.125$). It is worth noting that for both
protocols, the amount of classical information disclosed during the
classical sifting phase is lower than that of the SARG04 protocol.
This reduction decreases the probability of an information gain by
Eve using the announced classical information. One can use either of our two protocols as needed
for the necessary task.
\end{document}